\documentclass[10pt,
twocolumn,
 aps,
 jmp,%
 showkeys,
 amsmath,amssymb,
reprint,%
]{revtex4-1}

\usepackage[breaklinks=true,colorlinks=true,linkcolor=blue,urlcolor=blue,citecolor=blue]{hyperref}
\usepackage{graphicx}%
\usepackage{graphics}%
\usepackage{epstopdf}
\usepackage{dcolumn}%
\usepackage{soul,color}
\usepackage{bm}

\begin{document}
\title[]{Interplay of flux guiding and Hall effect in Nb films with nanogrooves}

\author{O.~V.~Dobrovolskiy$^{1,2}$, M.~Hanefeld$^1$, M.~Z\"orb$^1$, M.~Huth$^1$, and V.~A.~Shklovskij$^2$}
\address{$^1$Physikalisches Institut, Goethe University, 60438 Frankfurt am Main, Germany}
\address{$^2$Physics Department, V. Karazin National University, 61077 Kharkiv, Ukraine}

\begin{abstract}
The interplay between vortex guiding and the Hall effect in superconducting Nb films with periodically arranged nanogrooves is studied via four-probe measurements in standard and Hall configurations and accompanying theoretical modeling. The nanogrooves are milled by focused ion beam and induce a symmetric pinning potential of the washboard type. The resistivity tensor of the films is determined in the limit of small current densities at temperatures close to the critical temperature for the fundamental matching configuration of the vortex lattice with respect to the pinning nanolandscape. The angle between  the current direction with respect to the grooves is set at seven fixed values between $0^\circ$ and $90^\circ$. A sign change is observed in the temperature dependence of the Hall resistivity $\rho_\perp^-$ of as-grown films in a narrow temperature range near $T_c$. By contrast, for all nanopatterned films $\rho_\perp^-$ is nonzero in a broader temperature range below $T_c$, allowing us to discriminate between two contributions in $\rho_\perp^-$, namely one contribution originating from the guided vortex motion and the other one caused by the Hall anomaly just as in as-grown Nb films. All four measured resistivity components are successfully fitted to analytical expressions derived within the framework of a stochastic model of competing isotropic and anisotropic pinning. This provides evidence of the model validity for the description of the resistive response of superconductor thin films with washboard pinning nanolandscapes.
\end{abstract}

\maketitle

\section{Introduction}
Mixed-state Hall effect measurements represent a widely used approach to study the nonlinear vortex dynamics in superconductors, both in flat and nanopatterned samples \cite{Fio68prl,Not76ssc,Hag90prb,Hag91prb,Ric92prb,Hag93prb,Kun94prl,Pro98pcs,Pas99prl,Gob00prb,Lan01pcs,Pui04prb,Wor04prb,Wor06pcs,Ric06prb,Pui09prb,Seg11prb,Wan11prb,Sat13prb}. When a sample contains a directional pinning anisotropy, in normal-to-film plane magnetic fields $H$ at not too strong current densities $j$ directed at an arbitrary tilt angle with respect to the anisotropy axis of the pinning potential, the direction of the vortex motion is deflected from the direction of the Lorentz force $\mathbf{F}_L = n(\Phi_0/c)\,\mathbf{j}\times\mathbf{z}$. Here $n = \pm1$, $\Phi_0$ is the magnetic flux quantum, $c$ is the speed of light, and $\mathbf{z}$ is the unit vector in the direction of $\mathbf{H}$. This non-collinearity of the average vortex velocity $\mathbf{v}$ with the direction of the Lorentz force $\mathbf{F}_L$ is well known as the vortex guiding effect, i.\,e. vortices move more easily along the pinning channels rather than overcome the pinning barriers. Guided vortex motion has been extensively studied in various pinning nanolandscapes \cite{Nie69jap,Sil02prb,Sil03pcs,Sil03prb,Vil03prb,Wor04prb,Wor04pcs,Cri05prb,Luk06jap,Gon07jap,Sor07prb,Dob12njp,Dob10sst,Dob11pcs}, see e.\,g. \cite{Sil10inb} for a recent review. Vortex guiding is a valuable ``tool'' for controlling the motion of vortices in Abrikosov fluxonics \cite{Dob15arx}.

\begin{figure}[b!]
    \centering
    \includegraphics[width=1\linewidth]{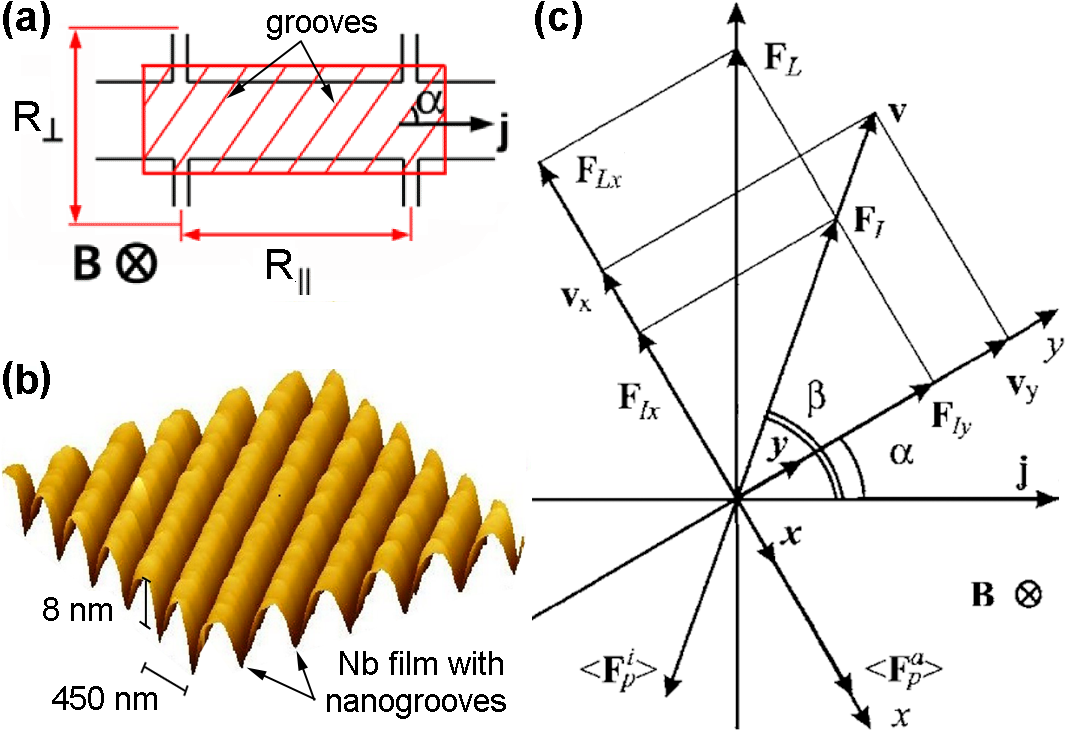}
    \caption{(a) Experimental geometry of the four-probe measurements in standard and Hall configurations. (b) Atomic force microscope image of the nanogroove array on the surface of the Nb film taken in non-contact mode. (c) Diagram of the forces acting on a vortex. The system of coordinates $xy$ with the unit vectors $\mathbf{x}$ and $\mathbf{y}$ is associated with the pinning channels which are parallel to the vector $\mathbf{y}$. The transport current density vector $\mathbf{j}$ is applied at an angle $\alpha$ with respect to $\mathbf{y}$. $\beta$ is the angle between the average velocity vector $\langle \mathbf{v}\rangle$ and $\mathbf{j}$ and it is a measure of the guiding effect. $\langle \mathbf{F}_p^a\rangle$ is the average pinning force induced by the washboard pinning landscape, $\langle \mathbf{F}_p^i\rangle$ is the average pinning force induced by the uncorrelated disorder, $\mathbf{F}_L$ is the Lorenz force for a vortex, $\mathbf{F}_I$ is the effective driving force, and $\mathbf{B}$ is the magnetic field vector.}
    \label{fGrooves}
\end{figure}

In consequence of vortex guiding, an even-in-field reversal transverse resistivity component $\rho^+_{\perp}$ appears \cite{Kop89etp,Shk99etp,Sil10inb}. It shows up in addition to the Hall resistivity component $\rho^-_{\perp}$ which is odd with regard to magnetic field reversal. A conventional measure for the vortex guiding effect is the \emph{guiding angle} $\beta$ between $\mathbf{v}$ and $\mathbf{j}$ \cite{Shk03ltp,Shk05ltp,Shk06prb}. For this angle an analytical expression was derived in the framework of a stochastic model of competing isotropic and anisotropic pinning \cite{Shk03ltp,Shk05ltp,Shk06prb}. This model allows one to describe the resistive response of a superconducting film with an anisotropic pinning landscape of the \emph{washboard} type in a perpendicular magnetic field as a function of temperature, current value, and transport current tilt angle $\alpha$ with respect to the pinning channels, refer to figure \ref{fGrooves}(a) for the geometry. Namely, the guiding angle $\beta$ is defined via $\cot\,\beta = - \rho^+_{\perp}/\rho^+_{\parallel}$, where $\rho^+_{\parallel}$ is the even-in-field longitudinal resistivity component. {Previously, an analytical account for vortex guiding in two-dimensional periodic pinning landscapes was given for the cases of incommensurate} \cite{Sil02prb} {and commensurate} \cite{Sil03pcs} {vortex lattice.} An analytical description of the vortex guiding effect in superconductors with anisotropic pinning was later on elaborated for pinning landscapes of the washboard type \cite{Shk99etp,Shk03ltp,Shk05ltp,Shk06prb}. At the same time, {a comparison of experimental data acquired on superconductors with washboard pinning nanostructures with the full set of derived expressions} \cite{Shk99etp,Shk03ltp,Shk05ltp,Shk06prb} has not been done so far.

It has also been pointed out \cite{Shk99etp,Shk03ltp,Shk05ltp,Shk06prb} and experimentally confirmed \cite{Pro98pcs,Sor07prb} that an additional longitudinal odd-in-field resistivity component $\rho^-_{\parallel}$ appears in consequence of the competing guided vortex motion and the Hall effect.
An analysis of the Hall resistivity $\rho^-_{\perp}$ is complicated by the presence of a puzzling feature of $\rho^-_{\perp}$ in the vicinity of the superconducting transition \cite{Not76ssc,Hag90prb,Hag91prb,Ric92prb,Hag93prb,Kun94prl,Pas99prl,Gob00prb,Lan01pcs,Pui04prb,Wor04prb,Wor06pcs,Ric06prb,Sor07prb,Pui09prb,Seg11prb,Wan11prb,Sat13prb}.
Namely, the Hall resistivity often changes its sign as a function of magnetic field and temperature; this is referred to as the \emph{anomalous} Hall effect. This anomaly of the mixed-state Hall effect was observed in non-patterned samples \cite{Seg11prb} as well as in samples with a directional pinning anisotropy \cite{Sor07prb}. While a number of mechanisms have been proposed to elucidate the nature of this anomaly, the physics behind this phenomenon remains debatable so far, see \cite{Bra95rpp} for a voluminous series of suggested mechanisms.

On the assumption that the ``bare'' Hall coefficient $\alpha_H$ is constant, several resistivity scaling laws have been derived theoretically for different pinning potentials. Vinokur \emph{et al.} \cite{Vin93prl} considered the momentum balance and proposed a model where a system of interacting vortices under quenched disorder and thermal noise follows a universal scaling law $\rho_\perp\propto\rho_\parallel^2$, which is a general feature of any isotropic vortex dynamics with an average pinning force directed along the average vortex-velocity vector. Wang \emph{et al.} \cite{Wan94prl} considered backflow currents and thermal fluctuations. They suggested that the scaling exponent varies from 2 to 1.5 as the pining strength increases. Shklovskij \emph{et al.} showed \cite{Shk99etp} for a washboard pinning potential that the form of the corresponding scaling relation is highly anisotropic since the pinning force for anisotropic pins is directed perpendicular to the pinning channels. The scaling law for the vortex motion across the potential channels reads $\rho_\perp\propto\rho_\parallel$ that can be interpreted as a scaling law with the exponent equal to unity. More complicated scaling relations were derived in Refs. \cite{Shk03ltp,Shk05ltp,Shk06prb} for the case of competing anisotropic and isotropic pinning. The reason for this is that the Hall resistivity can be represented as a sum of three terms with different signs, see the last line in expression (\ref{eAllR}). This is distinct from the Hall conductivity terms since the off-diagonal components of the conductivity tensor are not influenced by the presence of the isotopic and anisotropic pins within the framework of this model \cite{Shk03ltp,Shk05ltp,Shk06prb}.

On the experimental side, the most extensive work was done to shed light on peculiarities of the Hall effect in high-temperature superconductors \cite{Hag90prb,Hag91prb,Ric92prb,Hag93prb,Kun94prl,Pas99prl,Gob00prb,Lan01pcs,Pui04prb,Wor04prb,Wor06pcs,Ric06prb,Pui09prb,Seg11prb,Wan11prb,Sat13prb}.
An anomaly of the Hall resistivity was also discussed for conventional superconductors such as Pb \cite{Seg11prb}, MgB$_2$ \cite{Vas04pcs}, and Nb \cite{Sor07prb}. In particular, self-organization was used \cite{Hut02afm,Sor07prb} to provide semi-periodic, linearly extended pinning ``sites'' by spontaneous facetting of m-plane sapphire substrate surfaces on which Nb films were grown. It was demonstrated \cite{Sor07prb} that a pronounced guiding of vortices occurs. The authors of Ref. \cite{Sor07prb} succeeded to fit their data for the longitudinal even-in-field resistivity component to the theoretical expressions derived in Ref. \cite{Shk06prb}. However, the non-availability of non-patterned reference samples did not allow the authors to discriminate the anomalous Hall contribution from the guiding-induced contribution to the odd transverse resistivity.

Recently, we studied the vortex guiding effect in Nb films with focused ion beam-milled nanogroove arrays \cite{Dob12njp}. A tunable guiding of vortices was observed when the location of vortices in the assumed triangular vortex lattice with lattice parameter $a_\bigtriangleup = (2\Phi_0/H\sqrt{3})^{1/2}$ was adjusted by fine-tuning the magnetic field value $H$ to fulfill the geometrical matching condition $a_\bigtriangleup = 2a/\sqrt{3}$, where $a$ is the period of the washboard nanolandscape. In this way either the weak background isotropic pinning between grooves or the strong anisotropic pinning at the groove bottoms was probed by vortices. Employing the theoretical model \cite{Shk06prb} for a saw-tooth washboard pinning potential we were able to quantify the activation energies of the anisotropic pinning at the groove bottoms and the isotropic pinning between them.

Here, using a series of similar samples we extend that study \cite{Dob12njp} to the full set of magneto-resistivity tensor components and discuss the interplay of the vortex guiding effect with the Hall effect in the mixed state by fitting the resistivity data to the analytical expressions derived within the framework of the stochastic model of competing isotropic and anisotropic pinning \cite{Shk03ltp,Shk05ltp,Shk06prb}.

\section{Experiment}
The samples are eight 6-contact bridges defined by conventional lithography in a 72\,nm-thick Nb film. The film was deposited by dc magnetron sputtering with a sputtering rate of $1$\,nm/s in an ultra-high vacuum setup. The substrate was an epitaxially-polished a-plane sapphire substrate kept at $850^\circ$C during the deposition process. Details of the morphology characterization of the as-grown film are given in \cite{Dob12tsf}, where the film is referred to as film C. Seven bridges were nanostructured with a 450\,nm-periodic pinning profile shown in figure \ref{fGrooves}(b) by focused ion beam milling. The parameters of the patterning process were the same as in Ref. \cite{Dob12njp}. In each bridge the grooves are tilted at an angle $\alpha$ of $0^\circ$, $15^\circ$, $30^\circ$, $45^\circ$, $60^\circ$, $75^\circ$, and $90^\circ$ with respect to the direction of the transport current, see figure \ref{fGrooves}(a) for geometry. One bridge was left without grooves for documenting the anomalous Hall effect in as-grown Nb films. The width of the voltage contacts was $10\,\mu$m, that is, a factor of twenty larger than the spatial extend of one nanostructure period.

The electrical resistance was measured via four-probe measurements in standard and Hall configurations under normal-to-film plane magnetic field reversal (field ``up'' and field ``down''). The even and odd resistivity components with respect to the field reversal were determined according to the standard relations
\begin{equation}
    \label{eEvenOdd}
    \rho^{\pm}_{\parallel,\perp} = [\rho_{\parallel,\perp}(+\mathbf{H}) \pm \rho_{\parallel,\perp}(-\mathbf{H})]/2.
\end{equation}

All measurements were done at $H=8.8$\,mT corresponding to the fundamental matching configuration of vortices in a 450\,nm-periodic washboard nanolandscape, as shown in the inset to figure \ref{fAllR}(a). At this field value all vortices are pinned at the groove bottoms and there are no interstitial vortices. For periodic pinning landscapes, like the one used here, it has been shown by computer simulations \cite{Luq07prb} that for the fundamental matching configuration the vortex-vortex interaction is effectively cancelled. Provided the vortices move coherently, the dynamics of the entire vortex ensemble can be regarded as that of the \emph{single average vortex} in a \emph{mean pinning potential}, thus allowing one to employ single-vortex models for analyzing the resistance data.

All samples are characterized by a superconducting transition temperature $T_c = 8.61$\,K. It was determined at the lowest temperature at which the guiding angle $\beta(T) =90^\circ$ for $\alpha\neq 0^\circ$ and $90^\circ$, see figure \ref{fAngles}(a). The thickness of the as-grown film ensured that $T_c$ of all nanopatterned bridges with a groove depth of $8\pm0.5$\,nm is equal to the transition temperature of the as-grown film within an error of $0.01$\,K. All samples are characterized by an upper critical field at zero temperature $H_{c2}(0)$ of about $1$\,T as deduced from fitting the dependence $H_{c2}(T)$ to the phenomenological law $H_{c2}(T) = H_{c2}(0) [1-(T/T_c)^2]$.

\section{Results}
\begin{figure*}[th!]
    \centering
    \includegraphics[width=0.49\linewidth]{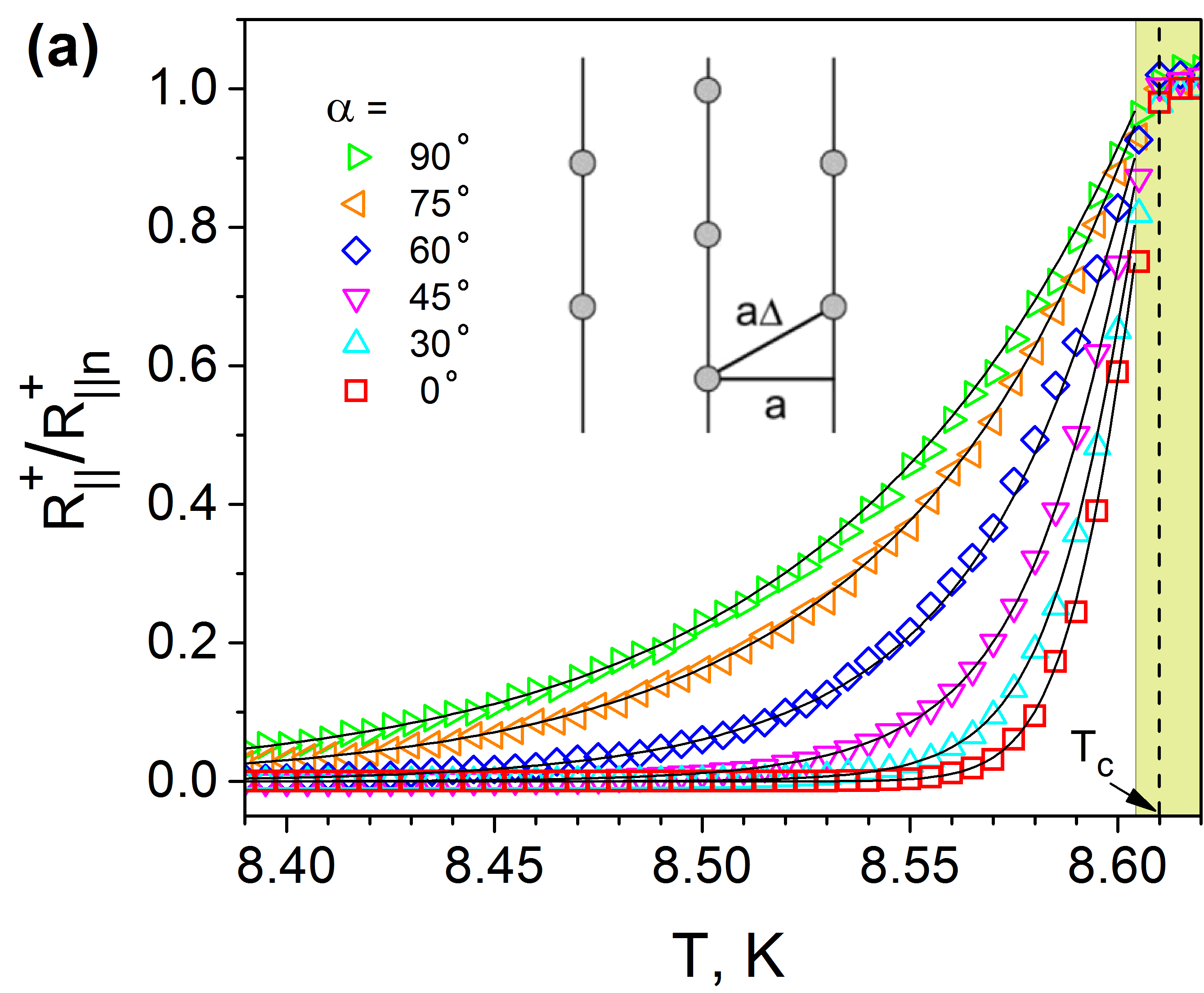}
    \includegraphics[width=0.49\linewidth]{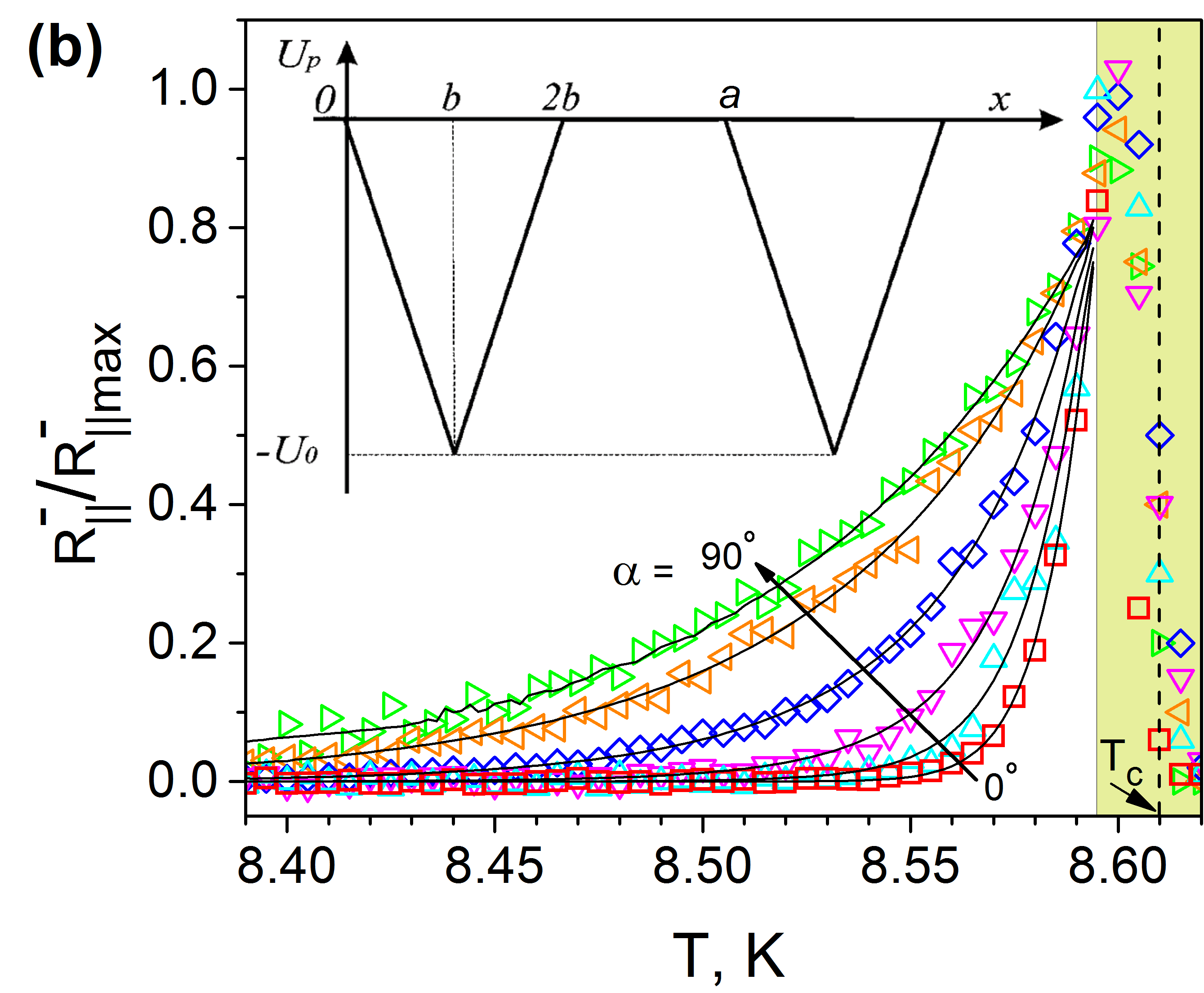}
    \includegraphics[width=0.49\linewidth]{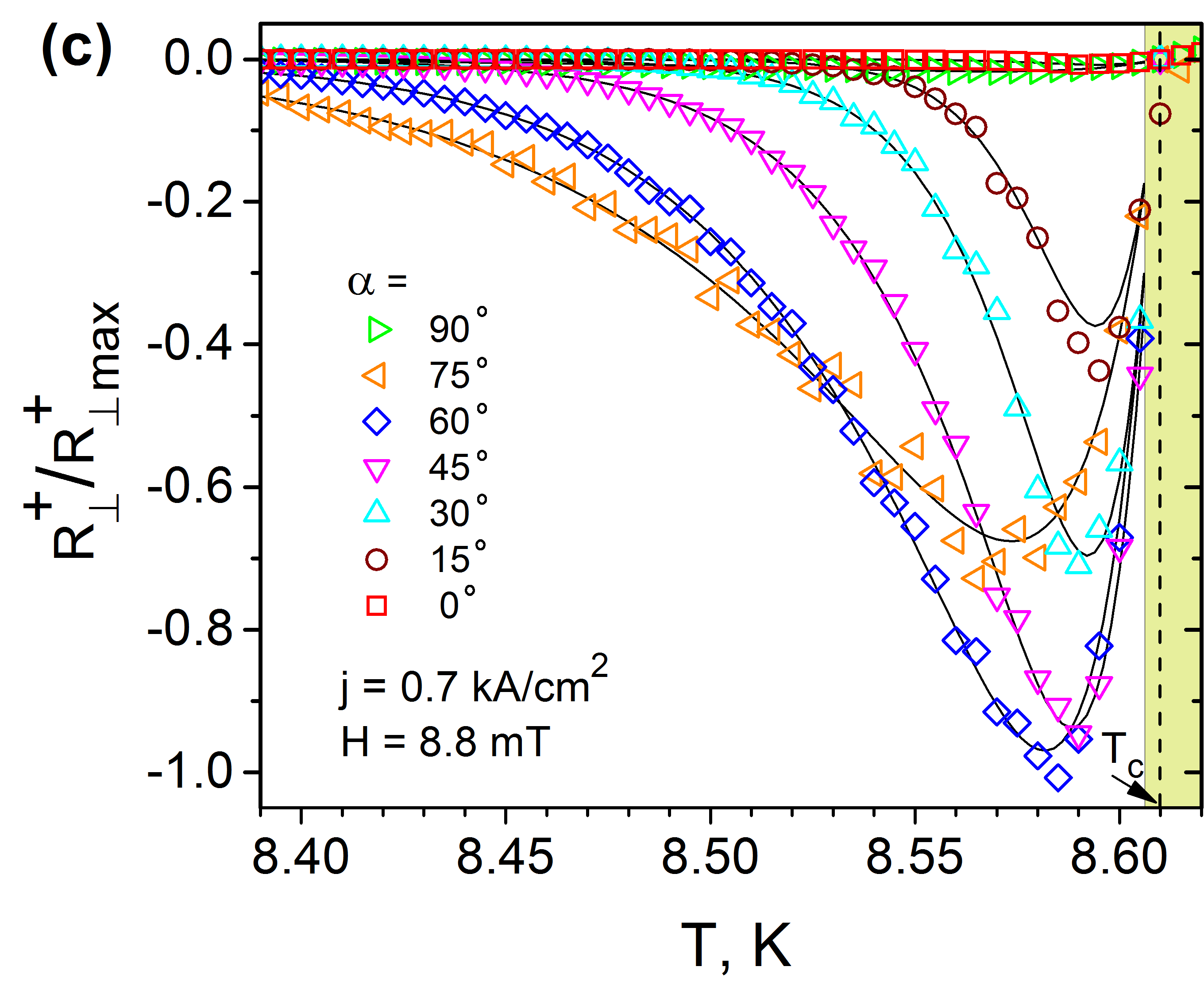}
    \includegraphics[width=0.49\linewidth]{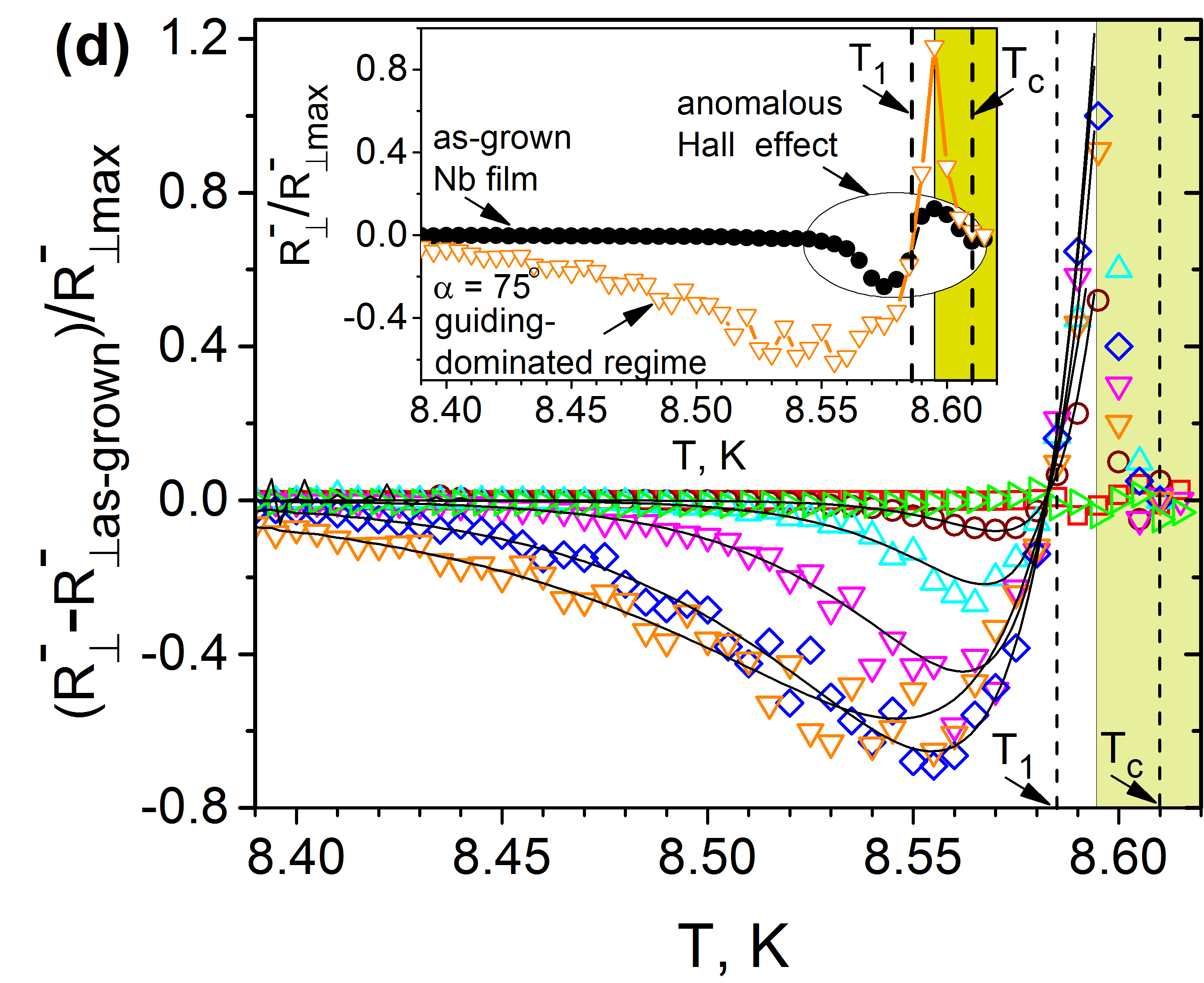}
    \caption{Temperature dependences of the normalized (a) even longitudinal, (b) odd longitudinal, (c) even transverse  and (d) odd transverse resistance components at the matching field $8.8$\,mT and the current density $0.7$\,kA/cm$^2$. Symbols: experiment, solid lines: fits to expressions (\ref{eAllR}) with the parameters detailed in the text. Inset in (a): The vortex lattice configuration with lattice parameter $a_\bigtriangleup = (2\Phi_0/H\sqrt{3})^{1/2}$ and the matching condition $a_\bigtriangleup = 2a/\sqrt{3}$ in the nanogroove array with the period $a=450$\,nm at $8.8$\,mT. Inset in (b): The saw-tooth pinning potential used in the theoretical modeling. Inset in (d): The raw $R^-_{\perp}(T)$ data for the sample with $\alpha = 75^\circ$ and the reference $R^-_{\perp}(T)_{\mathrm{as-grown}}$ Hall component in the as-grown Nb film. The shaded vertical bars in all panels mark the temperature range in which the resistance components can no longer be fitted by expressions (\ref{eAllR}).}
    \label{fAllR}
\end{figure*}
The temperature dependences of the four resistance components are shown in figure \ref{fAllR}. The even longitudinal resistance is normalized by the normal-state resistance value, while the other components are scaled by their respective maximum values. The acquired data demonstrate a systematic anisotropic character. Namely, the even longitudinal resistance component $R^+_{\parallel}(T)$ in figure \ref{fAllR}(a) exhibits smoother slopes with increasing $\alpha$ such that a higher resistive state persists down to lower temperatures. This is caused by the vortex guiding effect which has been extensively discussed, e.\,g. in Ref. \cite{Dob12njp}. The odd longitudinal resistance component $R^-_{\parallel}(T)$ in figure \ref{fAllR}(b) is characterized by a maximum at the position of the most steep ascend of the $R^+_{\parallel}(T)$ curves. On the low-temperature side of the maximum one recognizes resistance ``tails'' reminiscent of the behavior of $R^+_{\parallel}(T)$. The even transverse resistance component $R^+_{\perp}(T)$ in figure \ref{fAllR}(c) is negative. It has a minimum for all angles $\alpha\neq0^\circ,90^\circ$ and vanishes upon reaching $T_c$. The deeper minima are attained at $\alpha = 45^\circ$ and $60^\circ$. The $R^+_{\perp}(T)$ curves for $\alpha = 0^\circ$ and $90^\circ$ are zero in the entire temperature range. This can be understood as a consequence of an absent guiding force component along the grooves in these limiting cases.

The behavior of the $R^-_{\perp}(T)$ component in figure \ref{fAllR}(d) is most complex. The raw data for $R^-_{\perp}(T)$ for $\alpha = 75^\circ$ are shown in the inset of figure \ref{fAllR}(d) in comparison with the $R^-_{\perp}(T)$ data for the non-patterned reference Nb film. The Hall signal is nonzero in a narrow temperature range near $T_c$. A sign change of the $R^-_{\perp}(T)$ component for the non-structured reference bridge is observed at $T_1\approx 8.585$\,K and it is attributed to the anomalous Hall effect in the mixed state. Since this feature is an intrinsic property of all as-grown samples, in the main panel of figure \ref{fAllR}(d) we present the data for $R^-_{\perp}(T)$ with the reference $R^-_{\perp}(T)_{\mathrm{as-grown}}$ data subtracted. At lower temperatures $T<T_1$ the resulting curves $R^-_{\perp}(T)$ for $\alpha\neq0^\circ$ and $90^\circ$ mimic the features of the component $R^+_{\perp}(T)$ having a minimum which is most pronounced for intermediate angles $\alpha$. At the temperature $T_1 \approx 8.585$\,K the component $R^-_{\perp}(T)$ changes its sign from negative to positive as the temperature increases. Between $T_1$ and $T_c$ the $R^-_{\perp}(T)$ curves exhibit a maximum which, again, is most pronounced for the intermediate angles $\alpha$.

Figure \ref{fAngles} displays the temperature dependences of the two complementary measures for the guiding effect, namely the guiding angle $\beta(T) = -\mathrm{arccot} \rho^+_{\perp}/\rho^+_{\parallel}$ and the electric field angle $\theta_E(T)$, which is defined by expression (\ref{eThetaE}) and characterizes the nonlinear anisotropic mobility of the vortices in the washboard nanolandscape. In figure \ref{fAngles}(a) one sees that $\beta\approx\alpha$ up to $T = T_{\mathrm{FG}} \approx 8.48$\,K. This means that the full guiding (FG) of vortices takes place in the temperature range $T\lesssim T_{\mathrm{FG}}$. With increasing temperature a nonlinear crossover to $\beta =90^\circ$ ensues. The experimental data have a more pronounced scattering at lower temperatures since the measured resistive responses, especially their odd components, become more noisy at low temperatures. Upon approaching the value $\beta =90^\circ$ the slope of $\beta(T)$ for $\alpha\neq0^\circ$ and $90^\circ$ is well defined. This allows us to use the lowest temperature where $\beta(T)=90^\circ$ for the definition of the superconducting transition temperature $T_c$ which will be used in numerical simulations next.


\section{Analysis and discussion}
{To interpret the resistivity data related to the dynamics of the whole vortex \emph{ensemble}, we make use a rather crude approximation that at the fundamental matching field the effective vortex-vortex interaction is cancelled and the vortex dynamics is \emph{coherent}. When the vortices move in the transverse pinning landscape at $\alpha = 0^\circ$, the coherence in the vortex motion becomes apparent in independent experiments with superimposed dc and ac stimuli, leading to the appearance of quantum interference effects (Shapiro steps) in the current-voltage curves of the samples} \cite{Dob15mst,Dob15snm} {and new effects in the microwave-driven vortex dynamics} \cite{Dob15apl,Dob15met,Sil17inb}. {When the field value is tuned away from the matching condition, Shapiro steps disappear due to the lack of coherence in the vortex motion. Although the assumption of the coherent motion of the vortex lattice in transverse pinning landscapes is supported by the results of computer simulations} \cite{Luq07prb} {in the limit of \emph{zero} random pinning and \emph{without} thermal fluctuations at $\alpha = 0^\circ$, this assumption is likely to be safe only for the limiting cases of vortices moving strictly along or across the guiding channels of the pinning potential. In the general case of arbitrary $\alpha$ the long-range order of the moving vortex lattice is not expected to hold} \cite{Kos94prl} {and further effects originating from the long-range order lack can not be accounted for. Still, being aware of its limitations, we employ a \emph{single-vortex} stochastic model of competing isotropic and anisotropic pinning} \cite{Shk06prb} {for analyzing the resistivity data at all $\alpha$}.

The modelling is performed on the basis of an exact \emph{analytical} solution \cite{Shk06prb} of the Langevin equation for a vortex moving with velocity $\mathbf{v}$ in a magnetic field $\mathbf{B}=\mathbf{n}B$ with $B\equiv|\mathbf{B}|$, $\mathbf{n}=n\mathbf{z}$, $\mathbf{z}$ being the unit vector in the $z$-direction and $n=\pm 1$, viz.
\begin{equation}
    \label{eLE}
    \eta_{0}\mathbf{v}+n\alpha_{H}\mathbf{v}\times\mathbf{z}=\mathbf{F}_{L}+\mathbf{F}_{p}^{a}+\mathbf{F}_{p}^{i}+\mathbf{F}_{th},
\end{equation}
where $\mathbf{F}_{L}=n(\Phi_{0}/c)\mathbf{j}\times\mathbf{z}$ is the Lorentz force, $\mathbf{F}_{p}^{a}=-\nabla U_{p}(x)$ is the anisotropic pinning force induced by the nanogroove array, $\mathbf{F}_{p}^{i}$ is the isotropic pinning force induced by uncorrelated point-like disorder due to structural imperfectness of the samples, $\mathbf{F}_{th}$ is the thermal fluctuation force, $\eta_{0}$ is the vortex viscosity, and $\alpha_{H}$ is the Hall constant. A diagram of the forces acting on a vortex is sketched in figure \ref{fGrooves}(c).

We use a model saw-tooth washboard pinning potential of the following form
\begin{equation}
    \label{eWPP}
    U_{p}(x)=
    \left\{
        \begin{array}{ccl}
            - F_{p}x, \qquad\qquad\, 0 \leqslant x \leqslant b,\\
            F_{p}(x-2b), \qquad  b  \leqslant x \leqslant 2b,\\
            0, \qquad\qquad\quad\,\,\,\,\,2b \leqslant x \leqslant a, \\
        \end{array}
    \right.
\end{equation}
where $F_p = U_0/b$ is the pinning force with $U_0$ being the depth and $2b$ the width of the potential well, while $a$ is the period. The model pinning potential given by expression (\ref{eWPP}) is shown in the inset to figure \ref{fAllR}(b).

Within the framework of the model of competing isotropic and anisotropic pinning \cite{Shk03ltp,Shk05ltp,Shk06prb} the following expressions were derived for the resistivity components
\begin{equation}
    \label{eAllR}
        \begin {array}{ll}
\rho_{\parallel}^+ = \rho_f[\sin^2\alpha + \nu_a\cos^2\alpha]\nu_i,\\[2mm]
\rho_{\parallel}^- = \rho_f[(\sin^2\alpha + \nu_a\cos^2\alpha)\nu_i^- + \nu_i^2\nu_a^-\cos^2\alpha],\\[2mm]
\rho_{\perp}^+ = -\rho_f\nu_i[1-\nu_a]\sin\alpha\cos\alpha,\\[2mm]
\rho_{\perp}^- = \rho_f\big\{\big(n\epsilon\nu_a\nu_i^{2}+ [\nu_a^-\nu_i - \nu_i^- (1-\nu_a)]\big)\sin\alpha\cos\alpha\big\}.\\
        \end{array}
\end{equation}
Here the subscripts ``a'' and ``i'' relate to the characteristics of the anisotropic and isotropic pinning, $\rho_f$ is the flux-flow resistivity, and $\epsilon\equiv\alpha_H/\eta_0$ is the dimensionless Hall constant \cite{Shk06err}. The $\nu$-functions have the physical meaning of the effective mobility of the vortex in the respective pinning potential. The last formula in expressions (\ref{eAllR}) contains three terms with different signs. This yields the possibility for $\rho_{\perp}^-$ to change its sign in some range of the driving parameters, as analyzed in detail in Ref. \cite{Shk06prb}. {It should also be noted that the term $\nu_a^-$ depends on $\epsilon$} \cite{Shk06prb} {that emphasizes the fact that both odd resistivity components arise due to the competing guided vortex motion and Hall effect.}
\begin{figure}
    \centering
    \includegraphics[width=0.9\linewidth]{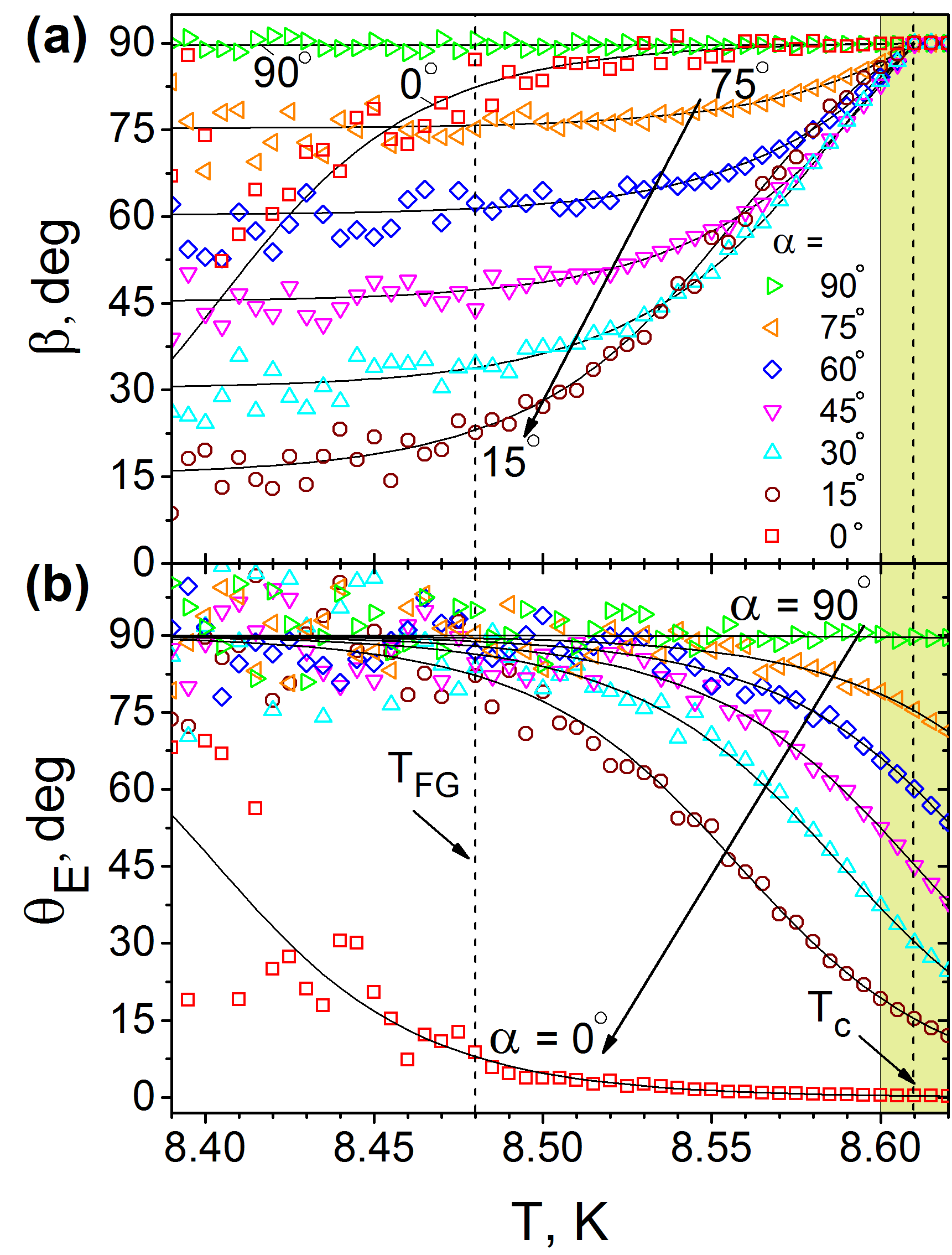}
    \caption{Temperature dependences of the guiding angle $\beta(T)$ (a) and the electric field angle $\theta_{E}$ (b) at the matching field $8.8$\,mT and the current density $0.7$\,kA/cm$^2$. Symbols: experiment, solid lines: fits to formulae (\ref{eCotB}) and (\ref{eThetaE}).}
    \label{fAngles}
\end{figure}

From these expressions it follows \cite{Shk06prb} that
\begin{equation}
    \label{eCotB}
    \cot\,\beta = - \frac{\rho^+_{\perp}}{\rho^+_{\parallel}} = \frac{1-\nu_a}{\tan\alpha + \nu_a\cot\alpha},
\end{equation}
where $\nu_a = \nu_a(j,T)$ is the effective mobility of the vortex in the anisotropic pinning potential under the action of the effective driving force $F_{Lx} = F_L\cos\alpha$. In expression (\ref{eCotB}) there is no term with subscript ``i''. Hence, the guiding angle $\beta$ is not influenced by the background isotropic pinning in the sample. This is due to the fact that its contributions are cancelled in the ratio (\ref{eCotB}) since isotropic pins only influence the \emph{magnitude} of the average velocity $\mathbf{v}$, but not its \emph{direction} \cite{Shk06prb}. {Although this conclusion follows within the framework of the single-vortex model used, it should be noted that an effective transverse pinning is known to emerge in the general case even if the pinning distribution is isotropic} \cite{Gia96prl}. {Turning back to Eq.} (\ref{eCotB}), the limiting case $\beta = \alpha$ corresponds to the full guiding of vortices, as opposed to the case $\beta = 90^\circ$ when the direction of the vortex motion coincides with the direction of the Lorentz force. The latter case is a consequence of isotropization of the resistive response due to the vanishing of the anisotropic pinning barriers under the action of transport current and temperature.

Employing expression (\ref{eCotB}) to the experimental data in figure \ref{fAngles}(a) it is possible to experimentally deduce the effective anisotropic vortex mobility $\nu_a(T)$ or, equivalently, the angle $\theta_E(T)$ between the electric field strength vector $\mathbf{E}$ and $\mathbf{j}$ \cite{Pas99prl,Shk06prb} being
\begin{equation}
    \label{eThetaE}
    \tan\theta_E (\alpha) = \tan\alpha/\nu_a.
\end{equation}
It is easy to see that $\theta_E = \alpha$ in the limiting case $\nu_a =1$ which corresponds to rather high temperatures and/or strong currents. In the opposite limiting case $\theta_E = 90^\circ$.


The resistivity tensor components (\ref{eAllR}) and expressions (\ref{eCotB}) and (\ref{eThetaE}) for the angles $\beta$ and $\theta_E$
were numerically evaluated as a function of temperature in the temperature range from $8$\,K to $8.61$\,K.
The calculated curves were fitted to the experimental data by variation of four free parameters: the volume fractions of the pinning centers between the grooves $\varepsilon_i$ and at the groove bottoms $\varepsilon_a$, as well as the average depth of the pinning potential well between the grooves $U_{0i}$ and at the groove bottoms $U_{0a}$. {While the characteristic lengths and energy scales of the material are known to depend strongly on temperature near $T_c$, two single temperature-independent effective values for each of the depths of the pinning potentials are used in the framework of the employed model.}

The best fits are shown by solid lines in figures \ref{fAllR} and \ref{fAngles}. The curves are calculated for $U_{0a}= 6300$\,K, $U_{0i} = 1944$\,K, $\varepsilon_a = 0.23$ and $\varepsilon_i = 0.1$. The definition of the parameter $\varepsilon = 2b/a$, where $a$ is the period of the pinning potential, is shown in the inset to figure \ref{fAllR}(b). The ratio $U_{0a}/U_{0i}$ is in line with the activation energies ratio quantified earlier for Nb films with anisotropic pinning \cite{Sor07prb,Dob12njp}. This points to the fact that the anisotropic pinning in the investigated samples dominates the isotropic one, as clearly seen in the temperature dependences of the four resistivity components.

The evaluated curves nicely reproduce the experimental data up to a temperature of about $8.6$\,K, within a fitting error less than $2\%$. In particular, expressions (\ref{eAllR}) approximate well the experimental data up to $T_{fit} = 8.605$\,K for the even resistivity components and up to $T_{fit} = 8.595$\,K for the odd resistivity components. Note that $T_1 < T_{fit} < T_c$, i.\,e., the fourth formula in expressions (\ref{eAllR}) describes well the sign reversal of the transverse odd resistivity component, after subtraction of the reference Hall signal in the as-grown film. Near $T_c$, the fits to expressions (\ref{eAllR}) are no longer possible since the transition to the normal state is not captured in the model of competing anisotropic and isotropic pinning \cite{Shk03ltp,Shk05ltp,Shk06prb}. A good agreement between the numerically evaluated expressions (\ref{eAllR}) and the experimental data also becomes apparent in figure \ref{fAngles} for the guiding angle $\beta$ and the electric field angle $\theta_E$. This provides evidence of the model validity for the description of the resistive response of superconductor thin films with washboard pinning nanolandscapes.

\section{Conclusion}
To summarize, the interplay between vortex guiding and the Hall effect in superconducting Nb films with periodically arranged nanogrooves has been studied via four-probe measurements in standard and Hall configurations and accompanying theoretical modeling. The temperature dependences of the four magnetoresistivity components of Nb films were measured for seven different current angle values with respect to the guiding direction of a washboard nanolandscape in Nb thin films under magnetic field reversal. All four resistivity components, the guiding angle $\beta$ and the electric field angle $\theta_E$ have been fitted well to the theoretical expressions derived within the framework of a stochastic model of competing anisotropic and isotropic pinning \cite{Shk03ltp,Shk05ltp,Shk06prb}, with one and the same set of fitting parameters. A sign change has been observed in the temperature dependence of the Hall resistivity $\rho_\perp^-$ of as-grown films in a narrow temperature range near $T_c$. By contrast, for all nanopatterned films $\rho_\perp^-$ is nonzero in a broader temperature range below $T_c$, that has allowed us to discriminate between two contributions in $\rho_\perp^-$, namely the contribution originating from the guided vortex motion in the washboard pinning nanolandscape and the other one caused by the Hall anomaly in as-grown Nb films. Although the sign change of the Hall resistivity in superconductors remains a puzzling problem having no ultimate explanation so far, we have shown that the guiding-induced contribution to $\rho_\perp^-$ can be approximated well by the theoretical expressions derived within the framework of the stochastic model of competing anisotropic and isotropic pinning. All in all, the reported results provide further evidence of the validity of the theoretical model \cite{Shk03ltp,Shk05ltp,Shk06prb} of competing anisotropic and isotropic pinning for the description of the resistive properties of superconductors with washboard pinning nanolandscapes.

\section*{Acknowledgements}
{\footnotesize
The authors thank R. Sachser for helping with nanopatterning and automating the data acquisition. This work was supported under the DFG project DO1511 and conducted within the framework of the COST Action MP1201 (NanoSC-COST) of the European Cooperation in Science and Technology.
}
\vspace{0.5cm}


\begin{thebibliography}{53}%
\makeatletter
\providecommand \@ifxundefined [1]{%
 \@ifx{#1\undefined}
}%
\providecommand \@ifnum [1]{%
 \ifnum #1\expandafter \@firstoftwo
 \else \expandafter \@secondoftwo
 \fi
}%
\providecommand \@ifx [1]{%
 \ifx #1\expandafter \@firstoftwo
 \else \expandafter \@secondoftwo
 \fi
}%
\providecommand \natexlab [1]{#1}%
\providecommand \enquote  [1]{``#1''}%
\providecommand \bibnamefont  [1]{#1}%
\providecommand \bibfnamefont [1]{#1}%
\providecommand \citenamefont [1]{#1}%
\providecommand \href@noop [0]{\@secondoftwo}%
\providecommand \href [0]{\begingroup \@sanitize@url \@href}%
\providecommand \@href[1]{\@@startlink{#1}\@@href}%
\providecommand \@@href[1]{\endgroup#1\@@endlink}%
\providecommand \@sanitize@url [0]{\catcode `\\12\catcode `\$12\catcode
  `\&12\catcode `\#12\catcode `\^12\catcode `\_12\catcode `\%12\relax}%
\providecommand \@@startlink[1]{}%
\providecommand \@@endlink[0]{}%
\providecommand \url  [0]{\begingroup\@sanitize@url \@url }%
\providecommand \@url [1]{\endgroup\@href {#1}{\urlprefix }}%
\providecommand \urlprefix  [0]{URL }%
\providecommand \Eprint [0]{\href }%
\providecommand \doibase [0]{http://dx.doi.org/}%
\providecommand \selectlanguage [0]{\@gobble}%
\providecommand \bibinfo  [0]{\@secondoftwo}%
\providecommand \bibfield  [0]{\@secondoftwo}%
\providecommand \translation [1]{[#1]}%
\providecommand \BibitemOpen [0]{}%
\providecommand \bibitemStop [0]{}%
\providecommand \bibitemNoStop [0]{.\EOS\space}%
\providecommand \EOS [0]{\spacefactor3000\relax}%
\providecommand \BibitemShut  [1]{\csname bibitem#1\endcsname}%
\let\auto@bib@innerbib\@empty
\bibitem [{\citenamefont {Fiory}\ and\ \citenamefont {Serin}(1968)}]{Fio68prl}%
  \BibitemOpen
  \bibfield  {author} {\bibinfo {author} {\bibfnamefont {A.~T.}\ \bibnamefont
  {Fiory}}\ and\ \bibinfo {author} {\bibfnamefont {B.}~\bibnamefont {Serin}},\
  }\href {\doibase 10.1103/PhysRevLett.21.359} {\bibfield  {journal} {\bibinfo
  {journal} {Phys. Rev. Lett.}\ }\textbf {\bibinfo {volume} {21}},\ \bibinfo
  {pages} {359} (\bibinfo {year} {1968})}\BibitemShut {NoStop}%
\bibitem [{\citenamefont {Noto}\ \emph {et~al.}(1976)\citenamefont {Noto},
  \citenamefont {Shinzawa},\ and\ \citenamefont {Muto}}]{Not76ssc}%
  \BibitemOpen
  \bibfield  {author} {\bibinfo {author} {\bibfnamefont {K.}~\bibnamefont
  {Noto}}, \bibinfo {author} {\bibfnamefont {S.}~\bibnamefont {Shinzawa}}, \
  and\ \bibinfo {author} {\bibfnamefont {Y.}~\bibnamefont {Muto}},\ }\href
  {\doibase http://dx.doi.org/10.1016/0038-1098(76)91245-X} {\bibfield
  {journal} {\bibinfo  {journal} {Solid State Commun.}\ }\textbf {\bibinfo
  {volume} {18}},\ \bibinfo {pages} {1081 } (\bibinfo {year}
  {1976})}\BibitemShut {NoStop}%
\bibitem [{\citenamefont {Hagen}\ \emph {et~al.}(1990)\citenamefont {Hagen},
  \citenamefont {Lobb}, \citenamefont {Greene}, \citenamefont {Forrester},\
  and\ \citenamefont {Kang}}]{Hag90prb}%
  \BibitemOpen
  \bibfield  {author} {\bibinfo {author} {\bibfnamefont {S.~J.}\ \bibnamefont
  {Hagen}}, \bibinfo {author} {\bibfnamefont {C.~J.}\ \bibnamefont {Lobb}},
  \bibinfo {author} {\bibfnamefont {R.~L.}\ \bibnamefont {Greene}}, \bibinfo
  {author} {\bibfnamefont {M.~G.}\ \bibnamefont {Forrester}}, \ and\ \bibinfo
  {author} {\bibfnamefont {J.~H.}\ \bibnamefont {Kang}},\ }\href {\doibase
  10.1103/PhysRevB.41.11630} {\bibfield  {journal} {\bibinfo  {journal} {Phys.
  Rev. B}\ }\textbf {\bibinfo {volume} {41}},\ \bibinfo {pages} {11630}
  (\bibinfo {year} {1990})}\BibitemShut {NoStop}%
\bibitem [{\citenamefont {Hagen}\ \emph {et~al.}(1991)\citenamefont {Hagen},
  \citenamefont {Lobb}, \citenamefont {Greene},\ and\ \citenamefont
  {Eddy}}]{Hag91prb}%
  \BibitemOpen
  \bibfield  {author} {\bibinfo {author} {\bibfnamefont {S.~J.}\ \bibnamefont
  {Hagen}}, \bibinfo {author} {\bibfnamefont {C.~J.}\ \bibnamefont {Lobb}},
  \bibinfo {author} {\bibfnamefont {R.~L.}\ \bibnamefont {Greene}}, \ and\
  \bibinfo {author} {\bibfnamefont {M.}~\bibnamefont {Eddy}},\ }\href {\doibase
  10.1103/PhysRevB.43.6246} {\bibfield  {journal} {\bibinfo  {journal} {Phys.
  Rev. B}\ }\textbf {\bibinfo {volume} {43}},\ \bibinfo {pages} {6246}
  (\bibinfo {year} {1991})}\BibitemShut {NoStop}%
\bibitem [{\citenamefont {Rice}\ \emph {et~al.}(1992)\citenamefont {Rice},
  \citenamefont {Rigakis}, \citenamefont {Ginsberg},\ and\ \citenamefont
  {Mochel}}]{Ric92prb}%
  \BibitemOpen
  \bibfield  {author} {\bibinfo {author} {\bibfnamefont {J.~P.}\ \bibnamefont
  {Rice}}, \bibinfo {author} {\bibfnamefont {N.}~\bibnamefont {Rigakis}},
  \bibinfo {author} {\bibfnamefont {D.~M.}\ \bibnamefont {Ginsberg}}, \ and\
  \bibinfo {author} {\bibfnamefont {J.~M.}\ \bibnamefont {Mochel}},\ }\href
  {\doibase 10.1103/PhysRevB.46.11050} {\bibfield  {journal} {\bibinfo
  {journal} {Phys. Rev. B}\ }\textbf {\bibinfo {volume} {46}},\ \bibinfo
  {pages} {11050} (\bibinfo {year} {1992})}\BibitemShut {NoStop}%
\bibitem [{\citenamefont {Hagen}\ \emph {et~al.}(1993)\citenamefont {Hagen},
  \citenamefont {Smith}, \citenamefont {Rajeswari}, \citenamefont {Peng},
  \citenamefont {Li}, \citenamefont {Greene}, \citenamefont {Mao},
  \citenamefont {Xi}, \citenamefont {Bhattacharya}, \citenamefont {Li},\ and\
  \citenamefont {Lobb}}]{Hag93prb}%
  \BibitemOpen
  \bibfield  {author} {\bibinfo {author} {\bibfnamefont {S.~J.}\ \bibnamefont
  {Hagen}}, \bibinfo {author} {\bibfnamefont {A.~W.}\ \bibnamefont {Smith}},
  \bibinfo {author} {\bibfnamefont {M.}~\bibnamefont {Rajeswari}}, \bibinfo
  {author} {\bibfnamefont {J.~L.}\ \bibnamefont {Peng}}, \bibinfo {author}
  {\bibfnamefont {Z.~Y.}\ \bibnamefont {Li}}, \bibinfo {author} {\bibfnamefont
  {R.~L.}\ \bibnamefont {Greene}}, \bibinfo {author} {\bibfnamefont {S.~N.}\
  \bibnamefont {Mao}}, \bibinfo {author} {\bibfnamefont {X.~X.}\ \bibnamefont
  {Xi}}, \bibinfo {author} {\bibfnamefont {S.}~\bibnamefont {Bhattacharya}},
  \bibinfo {author} {\bibfnamefont {Q.}~\bibnamefont {Li}}, \ and\ \bibinfo
  {author} {\bibfnamefont {C.~J.}\ \bibnamefont {Lobb}},\ }\href {\doibase
  10.1103/PhysRevB.47.1064} {\bibfield  {journal} {\bibinfo  {journal} {Phys.
  Rev. B}\ }\textbf {\bibinfo {volume} {47}},\ \bibinfo {pages} {1064}
  (\bibinfo {year} {1993})}\BibitemShut {NoStop}%
\bibitem [{\citenamefont {Kunchur}\ \emph {et~al.}(1994)\citenamefont
  {Kunchur}, \citenamefont {Christen}, \citenamefont {Klabunde},\ and\
  \citenamefont {Phillips}}]{Kun94prl}%
  \BibitemOpen
  \bibfield  {author} {\bibinfo {author} {\bibfnamefont {M.~N.}\ \bibnamefont
  {Kunchur}}, \bibinfo {author} {\bibfnamefont {D.~K.}\ \bibnamefont
  {Christen}}, \bibinfo {author} {\bibfnamefont {C.~E.}\ \bibnamefont
  {Klabunde}}, \ and\ \bibinfo {author} {\bibfnamefont {J.~M.}\ \bibnamefont
  {Phillips}},\ }\href {\doibase 10.1103/PhysRevLett.72.752} {\bibfield
  {journal} {\bibinfo  {journal} {Phys. Rev. Lett.}\ }\textbf {\bibinfo
  {volume} {72}},\ \bibinfo {pages} {752} (\bibinfo {year} {1994})}\BibitemShut
  {NoStop}%
\bibitem [{\citenamefont {Prodan}\ \emph {et~al.}(1998)\citenamefont {Prodan},
  \citenamefont {Shklovskij}, \citenamefont {Chabanenko}, \citenamefont
  {Bondarenko}, \citenamefont {Obolenskii}, \citenamefont {Szymczak},\ and\
  \citenamefont {Piechota}}]{Pro98pcs}%
  \BibitemOpen
  \bibfield  {author} {\bibinfo {author} {\bibfnamefont {A.~A.}\ \bibnamefont
  {Prodan}}, \bibinfo {author} {\bibfnamefont {V.~A.}\ \bibnamefont
  {Shklovskij}}, \bibinfo {author} {\bibfnamefont {V.~V.}\ \bibnamefont
  {Chabanenko}}, \bibinfo {author} {\bibfnamefont {A.~V.}\ \bibnamefont
  {Bondarenko}}, \bibinfo {author} {\bibfnamefont {M.~A.}\ \bibnamefont
  {Obolenskii}}, \bibinfo {author} {\bibfnamefont {H.}~\bibnamefont
  {Szymczak}}, \ and\ \bibinfo {author} {\bibfnamefont {S.}~\bibnamefont
  {Piechota}},\ }\href {\doibase
  http://dx.doi.org/10.1016/S0921-4534(98)00214-7} {\bibfield  {journal}
  {\bibinfo  {journal} {Physica C}\ }\textbf {\bibinfo {volume} {302}},\
  \bibinfo {pages} {271 } (\bibinfo {year} {1998})}\BibitemShut {NoStop}%
\bibitem [{\citenamefont {Pastoriza}\ \emph {et~al.}(1999)\citenamefont
  {Pastoriza}, \citenamefont {Candia},\ and\ \citenamefont {Nieva}}]{Pas99prl}%
  \BibitemOpen
  \bibfield  {author} {\bibinfo {author} {\bibfnamefont {H.}~\bibnamefont
  {Pastoriza}}, \bibinfo {author} {\bibfnamefont {S.}~\bibnamefont {Candia}}, \
  and\ \bibinfo {author} {\bibfnamefont {G.}~\bibnamefont {Nieva}},\ }\href
  {\doibase 10.1103/PhysRevLett.83.1026} {\bibfield  {journal} {\bibinfo
  {journal} {Phys. Rev. Lett.}\ }\textbf {\bibinfo {volume} {83}},\ \bibinfo
  {pages} {1026} (\bibinfo {year} {1999})}\BibitemShut {NoStop}%
\bibitem [{\citenamefont {G\"ob}\ \emph {et~al.}(2000)\citenamefont {G\"ob},
  \citenamefont {Liebich}, \citenamefont {Lang}, \citenamefont {Puica},
  \citenamefont {Sobolewski}, \citenamefont {R\"ossler}, \citenamefont
  {Pedarnig},\ and\ \citenamefont {B\"auerle}}]{Gob00prb}%
  \BibitemOpen
  \bibfield  {author} {\bibinfo {author} {\bibfnamefont {W.}~\bibnamefont
  {G\"ob}}, \bibinfo {author} {\bibfnamefont {W.}~\bibnamefont {Liebich}},
  \bibinfo {author} {\bibfnamefont {W.}~\bibnamefont {Lang}}, \bibinfo {author}
  {\bibfnamefont {I.}~\bibnamefont {Puica}}, \bibinfo {author} {\bibfnamefont
  {R.}~\bibnamefont {Sobolewski}}, \bibinfo {author} {\bibfnamefont
  {R.}~\bibnamefont {R\"ossler}}, \bibinfo {author} {\bibfnamefont {J.~D.}\
  \bibnamefont {Pedarnig}}, \ and\ \bibinfo {author} {\bibfnamefont
  {D.}~\bibnamefont {B\"auerle}},\ }\href {\doibase 10.1103/PhysRevB.62.9780}
  {\bibfield  {journal} {\bibinfo  {journal} {Phys. Rev. B}\ }\textbf {\bibinfo
  {volume} {62}},\ \bibinfo {pages} {9780} (\bibinfo {year}
  {2000})}\BibitemShut {NoStop}%
\bibitem [{\citenamefont {Lang}\ \emph {et~al.}(2001)\citenamefont {Lang},
  \citenamefont {G\"ob}, \citenamefont {Pedarnig}, \citenamefont {R\"ossler},\
  and\ \citenamefont {B\"auerle}}]{Lan01pcs}%
  \BibitemOpen
  \bibfield  {author} {\bibinfo {author} {\bibfnamefont {W.}~\bibnamefont
  {Lang}}, \bibinfo {author} {\bibfnamefont {W.}~\bibnamefont {G\"ob}},
  \bibinfo {author} {\bibfnamefont {J.~D.}\ \bibnamefont {Pedarnig}}, \bibinfo
  {author} {\bibfnamefont {R.}~\bibnamefont {R\"ossler}}, \ and\ \bibinfo
  {author} {\bibfnamefont {D.}~\bibnamefont {B\"auerle}},\ }\href {\doibase
  10.1016/S0921-4534(01)00841-3} {\bibfield  {journal} {\bibinfo  {journal}
  {Physica C}\ }\textbf {\bibinfo {volume} {364--365}},\ \bibinfo {pages} {518}
  (\bibinfo {year} {2001})}\BibitemShut {NoStop}%
\bibitem [{\citenamefont {Puica}\ \emph {et~al.}(2004)\citenamefont {Puica},
  \citenamefont {Lang}, \citenamefont {G\"ob},\ and\ \citenamefont
  {Sobolewski}}]{Pui04prb}%
  \BibitemOpen
  \bibfield  {author} {\bibinfo {author} {\bibfnamefont {I.}~\bibnamefont
  {Puica}}, \bibinfo {author} {\bibfnamefont {W.}~\bibnamefont {Lang}},
  \bibinfo {author} {\bibfnamefont {W.}~\bibnamefont {G\"ob}}, \ and\ \bibinfo
  {author} {\bibfnamefont {R.}~\bibnamefont {Sobolewski}},\ }\href {\doibase
  10.1103/PhysRevB.69.104513} {\bibfield  {journal} {\bibinfo  {journal} {Phys.
  Rev. B}\ }\textbf {\bibinfo {volume} {69}},\ \bibinfo {pages} {104513}
  (\bibinfo {year} {2004})}\BibitemShut {NoStop}%
\bibitem [{\citenamefont {W\"ordenweber}\ \emph {et~al.}(2004)\citenamefont
  {W\"ordenweber}, \citenamefont {Dymashevski},\ and\ \citenamefont
  {Misko}}]{Wor04prb}%
  \BibitemOpen
  \bibfield  {author} {\bibinfo {author} {\bibfnamefont {R.}~\bibnamefont
  {W\"ordenweber}}, \bibinfo {author} {\bibfnamefont {P.}~\bibnamefont
  {Dymashevski}}, \ and\ \bibinfo {author} {\bibfnamefont {V.~R.}\ \bibnamefont
  {Misko}},\ }\href {\doibase 10.1103/PhysRevB.69.184504} {\bibfield  {journal}
  {\bibinfo  {journal} {Phys. Rev. B}\ }\textbf {\bibinfo {volume} {69}},\
  \bibinfo {pages} {184504} (\bibinfo {year} {2004})}\BibitemShut {NoStop}%
\bibitem [{\citenamefont {W\"ordenweber}\ \emph {et~al.}(2006)\citenamefont
  {W\"ordenweber}, \citenamefont {Sankarraj}, \citenamefont {Dymashevski},\
  and\ \citenamefont {Hollmann}}]{Wor06pcs}%
  \BibitemOpen
  \bibfield  {author} {\bibinfo {author} {\bibfnamefont {R.}~\bibnamefont
  {W\"ordenweber}}, \bibinfo {author} {\bibfnamefont {J.}~\bibnamefont
  {Sankarraj}}, \bibinfo {author} {\bibfnamefont {P.}~\bibnamefont
  {Dymashevski}}, \ and\ \bibinfo {author} {\bibfnamefont {E.}~\bibnamefont
  {Hollmann}},\ }\href {\doibase 10.1016/j.physc.2005.11.014} {\bibfield
  {journal} {\bibinfo  {journal} {Physica C}\ }\textbf {\bibinfo {volume}
  {434}},\ \bibinfo {pages} {101} (\bibinfo {year} {2006})}\BibitemShut
  {NoStop}%
\bibitem [{\citenamefont {Richter}\ \emph {et~al.}(2006)\citenamefont
  {Richter}, \citenamefont {Puica}, \citenamefont {Lang}, \citenamefont
  {Peruzzi}, \citenamefont {Durrell}, \citenamefont {Sturm}, \citenamefont
  {Pedarnig},\ and\ \citenamefont {B\"auerle}}]{Ric06prb}%
  \BibitemOpen
  \bibfield  {author} {\bibinfo {author} {\bibfnamefont {H.}~\bibnamefont
  {Richter}}, \bibinfo {author} {\bibfnamefont {I.}~\bibnamefont {Puica}},
  \bibinfo {author} {\bibfnamefont {W.}~\bibnamefont {Lang}}, \bibinfo {author}
  {\bibfnamefont {M.}~\bibnamefont {Peruzzi}}, \bibinfo {author} {\bibfnamefont
  {J.~H.}\ \bibnamefont {Durrell}}, \bibinfo {author} {\bibfnamefont
  {H.}~\bibnamefont {Sturm}}, \bibinfo {author} {\bibfnamefont {J.~D.}\
  \bibnamefont {Pedarnig}}, \ and\ \bibinfo {author} {\bibfnamefont
  {D.}~\bibnamefont {B\"auerle}},\ }\href {\doibase 10.1103/PhysRevB.73.184506}
  {\bibfield  {journal} {\bibinfo  {journal} {Phys. Rev. B}\ }\textbf {\bibinfo
  {volume} {73}},\ \bibinfo {pages} {184506} (\bibinfo {year}
  {2006})}\BibitemShut {NoStop}%
\bibitem [{\citenamefont {Puica}\ \emph {et~al.}(2009)\citenamefont {Puica},
  \citenamefont {Lang}, \citenamefont {Siraj}, \citenamefont {Pedarnig},\ and\
  \citenamefont {B\"auerle}}]{Pui09prb}%
  \BibitemOpen
  \bibfield  {author} {\bibinfo {author} {\bibfnamefont {I.}~\bibnamefont
  {Puica}}, \bibinfo {author} {\bibfnamefont {W.}~\bibnamefont {Lang}},
  \bibinfo {author} {\bibfnamefont {K.}~\bibnamefont {Siraj}}, \bibinfo
  {author} {\bibfnamefont {J.~D.}\ \bibnamefont {Pedarnig}}, \ and\ \bibinfo
  {author} {\bibfnamefont {D.}~\bibnamefont {B\"auerle}},\ }\href {\doibase
  10.1103/PhysRevB.79.094522} {\bibfield  {journal} {\bibinfo  {journal} {Phys.
  Rev. B}\ }\textbf {\bibinfo {volume} {79}},\ \bibinfo {pages} {094522}
  (\bibinfo {year} {2009})}\BibitemShut {NoStop}%
\bibitem [{\citenamefont {Segal}\ \emph {et~al.}(2011)\citenamefont {Segal},
  \citenamefont {Karpovski},\ and\ \citenamefont {Gerber}}]{Seg11prb}%
  \BibitemOpen
  \bibfield  {author} {\bibinfo {author} {\bibfnamefont {A.}~\bibnamefont
  {Segal}}, \bibinfo {author} {\bibfnamefont {M.}~\bibnamefont {Karpovski}}, \
  and\ \bibinfo {author} {\bibfnamefont {A.}~\bibnamefont {Gerber}},\ }\href
  {\doibase 10.1103/PhysRevB.83.094531} {\bibfield  {journal} {\bibinfo
  {journal} {Phys. Rev. B}\ }\textbf {\bibinfo {volume} {83}},\ \bibinfo
  {pages} {094531} (\bibinfo {year} {2011})}\BibitemShut {NoStop}%
\bibitem [{\citenamefont {Wang}\ \emph {et~al.}(2011)\citenamefont {Wang},
  \citenamefont {Sou}, \citenamefont {Yang}, \citenamefont {Chang},
  \citenamefont {Cheng}, \citenamefont {Tsuei}, \citenamefont {Su},
  \citenamefont {Wolf},\ and\ \citenamefont {Adelmann}}]{Wan11prb}%
  \BibitemOpen
  \bibfield  {author} {\bibinfo {author} {\bibfnamefont {L.~M.}\ \bibnamefont
  {Wang}}, \bibinfo {author} {\bibfnamefont {U.-C.}\ \bibnamefont {Sou}},
  \bibinfo {author} {\bibfnamefont {H.~C.}\ \bibnamefont {Yang}}, \bibinfo
  {author} {\bibfnamefont {L.~J.}\ \bibnamefont {Chang}}, \bibinfo {author}
  {\bibfnamefont {C.-M.}\ \bibnamefont {Cheng}}, \bibinfo {author}
  {\bibfnamefont {K.-D.}\ \bibnamefont {Tsuei}}, \bibinfo {author}
  {\bibfnamefont {Y.}~\bibnamefont {Su}}, \bibinfo {author} {\bibfnamefont
  {T.}~\bibnamefont {Wolf}}, \ and\ \bibinfo {author} {\bibfnamefont
  {P.}~\bibnamefont {Adelmann}},\ }\href {\doibase 10.1103/PhysRevB.83.134506}
  {\bibfield  {journal} {\bibinfo  {journal} {Phys. Rev. B}\ }\textbf {\bibinfo
  {volume} {83}},\ \bibinfo {pages} {134506} (\bibinfo {year}
  {2011})}\BibitemShut {NoStop}%
\bibitem [{\citenamefont {Sato}\ \emph {et~al.}(2013)\citenamefont {Sato},
  \citenamefont {Katase}, \citenamefont {Kang}, \citenamefont {Hiramatsu},
  \citenamefont {Kamiya},\ and\ \citenamefont {Hosono}}]{Sat13prb}%
  \BibitemOpen
  \bibfield  {author} {\bibinfo {author} {\bibfnamefont {H.}~\bibnamefont
  {Sato}}, \bibinfo {author} {\bibfnamefont {T.}~\bibnamefont {Katase}},
  \bibinfo {author} {\bibfnamefont {W.~N.}\ \bibnamefont {Kang}}, \bibinfo
  {author} {\bibfnamefont {H.}~\bibnamefont {Hiramatsu}}, \bibinfo {author}
  {\bibfnamefont {T.}~\bibnamefont {Kamiya}}, \ and\ \bibinfo {author}
  {\bibfnamefont {H.}~\bibnamefont {Hosono}},\ }\href {\doibase
  10.1103/PhysRevB.87.064504} {\bibfield  {journal} {\bibinfo  {journal} {Phys.
  Rev. B}\ }\textbf {\bibinfo {volume} {87}},\ \bibinfo {pages} {064504}
  (\bibinfo {year} {2013})}\BibitemShut {NoStop}%
\bibitem [{\citenamefont {Niessen}\ and\ \citenamefont
  {Weijsenfeld}(1969)}]{Nie69jap}%
  \BibitemOpen
  \bibfield  {author} {\bibinfo {author} {\bibfnamefont {A.~K.}\ \bibnamefont
  {Niessen}}\ and\ \bibinfo {author} {\bibfnamefont {C.~H.}\ \bibnamefont
  {Weijsenfeld}},\ }\href {\doibase 10.1063/1.1657066} {\bibfield  {journal}
  {\bibinfo  {journal} {J. Appl. Phys.}\ }\textbf {\bibinfo {volume} {40}},\
  \bibinfo {pages} {384} (\bibinfo {year} {1969})}\BibitemShut {NoStop}%
\bibitem [{\citenamefont {de~Souza~Silva}\ and\ \citenamefont
  {Carneiro}(2002)}]{Sil02prb}%
  \BibitemOpen
  \bibfield  {author} {\bibinfo {author} {\bibfnamefont {C.~C.}\ \bibnamefont
  {de~Souza~Silva}}\ and\ \bibinfo {author} {\bibfnamefont {G.}~\bibnamefont
  {Carneiro}},\ }\href {\doibase 10.1103/PhysRevB.66.054514} {\bibfield
  {journal} {\bibinfo  {journal} {Phys. Rev. B}\ }\textbf {\bibinfo {volume}
  {66}},\ \bibinfo {pages} {054514} (\bibinfo {year} {2002})}\BibitemShut
  {NoStop}%
\bibitem [{\citenamefont {de~Souza~Silva}\ and\ \citenamefont
  {Carneiro}(2003)}]{Sil03pcs}%
  \BibitemOpen
  \bibfield  {author} {\bibinfo {author} {\bibfnamefont {C.~C.}\ \bibnamefont
  {de~Souza~Silva}}\ and\ \bibinfo {author} {\bibfnamefont {G.}~\bibnamefont
  {Carneiro}},\ }\href {\doibase
  http://dx.doi.org/10.1016/S0921-4534(03)00892-X} {\bibfield  {journal}
  {\bibinfo  {journal} {Physica C}\ }\textbf {\bibinfo {volume} {391}},\
  \bibinfo {pages} {203 } (\bibinfo {year} {2003})}\BibitemShut {NoStop}%
\bibitem [{\citenamefont {Silhanek}\ \emph {et~al.}(2003)\citenamefont
  {Silhanek}, \citenamefont {Van~Look}, \citenamefont {Raedts}, \citenamefont
  {Jonckheere},\ and\ \citenamefont {Moshchalkov}}]{Sil03prb}%
  \BibitemOpen
  \bibfield  {author} {\bibinfo {author} {\bibfnamefont {A.~V.}\ \bibnamefont
  {Silhanek}}, \bibinfo {author} {\bibfnamefont {L.}~\bibnamefont {Van~Look}},
  \bibinfo {author} {\bibfnamefont {S.}~\bibnamefont {Raedts}}, \bibinfo
  {author} {\bibfnamefont {R.}~\bibnamefont {Jonckheere}}, \ and\ \bibinfo
  {author} {\bibfnamefont {V.~V.}\ \bibnamefont {Moshchalkov}},\ }\href
  {\doibase 10.1103/PhysRevB.68.214504} {\bibfield  {journal} {\bibinfo
  {journal} {Phys. Rev. B}\ }\textbf {\bibinfo {volume} {68}},\ \bibinfo
  {pages} {214504} (\bibinfo {year} {2003})}\BibitemShut {NoStop}%
\bibitem [{\citenamefont {Villegas}\ \emph {et~al.}(2003)\citenamefont
  {Villegas}, \citenamefont {Gonzalez}, \citenamefont {Montero}, \citenamefont
  {Schuller},\ and\ \citenamefont {Vicent}}]{Vil03prb}%
  \BibitemOpen
  \bibfield  {author} {\bibinfo {author} {\bibfnamefont {J.~E.}\ \bibnamefont
  {Villegas}}, \bibinfo {author} {\bibfnamefont {E.~M.}\ \bibnamefont
  {Gonzalez}}, \bibinfo {author} {\bibfnamefont {M.~I.}\ \bibnamefont
  {Montero}}, \bibinfo {author} {\bibfnamefont {I.~K.}\ \bibnamefont
  {Schuller}}, \ and\ \bibinfo {author} {\bibfnamefont {J.~L.}\ \bibnamefont
  {Vicent}},\ }\href {\doibase 10.1103/PhysRevB.68.224504} {\bibfield
  {journal} {\bibinfo  {journal} {Phys. Rev. B}\ }\textbf {\bibinfo {volume}
  {68}},\ \bibinfo {pages} {224504} (\bibinfo {year} {2003})}\BibitemShut
  {NoStop}%
\bibitem [{\citenamefont {W\"ordenweber}\ and\ \citenamefont
  {Dymashevski}(2004)}]{Wor04pcs}%
  \BibitemOpen
  \bibfield  {author} {\bibinfo {author} {\bibfnamefont {R.}~\bibnamefont
  {W\"ordenweber}}\ and\ \bibinfo {author} {\bibfnamefont {P.}~\bibnamefont
  {Dymashevski}},\ }\href {\doibase 10.1016/j.physc.2003.11.024} {\bibfield
  {journal} {\bibinfo  {journal} {Physica C: Superconductivity}\ }\textbf
  {\bibinfo {volume} {404}},\ \bibinfo {pages} {421 } (\bibinfo {year}
  {2004})}\BibitemShut {NoStop}%
\bibitem [{\citenamefont {Crisan}\ \emph {et~al.}(2005)\citenamefont {Crisan},
  \citenamefont {Pross}, \citenamefont {Cole}, \citenamefont {Bending},
  \citenamefont {W\"ordenweber}, \citenamefont {Lahl},\ and\ \citenamefont
  {Brandt}}]{Cri05prb}%
  \BibitemOpen
  \bibfield  {author} {\bibinfo {author} {\bibfnamefont {A.}~\bibnamefont
  {Crisan}}, \bibinfo {author} {\bibfnamefont {A.}~\bibnamefont {Pross}},
  \bibinfo {author} {\bibfnamefont {D.}~\bibnamefont {Cole}}, \bibinfo {author}
  {\bibfnamefont {S.~J.}\ \bibnamefont {Bending}}, \bibinfo {author}
  {\bibfnamefont {R.}~\bibnamefont {W\"ordenweber}}, \bibinfo {author}
  {\bibfnamefont {P.}~\bibnamefont {Lahl}}, \ and\ \bibinfo {author}
  {\bibfnamefont {E.~H.}\ \bibnamefont {Brandt}},\ }\href {\doibase
  10.1103/PhysRevB.71.144504} {\bibfield  {journal} {\bibinfo  {journal} {Phys.
  Rev. B}\ }\textbf {\bibinfo {volume} {71}},\ \bibinfo {pages} {144504}
  (\bibinfo {year} {2005})}\BibitemShut {NoStop}%
\bibitem [{\citenamefont {Lukashenko}\ \emph {et~al.}(2006)\citenamefont
  {Lukashenko}, \citenamefont {Ustinov}, \citenamefont {Zhuravel},
  \citenamefont {Hollmann},\ and\ \citenamefont {W\"ordenweber}}]{Luk06jap}%
  \BibitemOpen
  \bibfield  {author} {\bibinfo {author} {\bibfnamefont {A.}~\bibnamefont
  {Lukashenko}}, \bibinfo {author} {\bibfnamefont {A.~V.}\ \bibnamefont
  {Ustinov}}, \bibinfo {author} {\bibfnamefont {A.~P.}\ \bibnamefont
  {Zhuravel}}, \bibinfo {author} {\bibfnamefont {E.}~\bibnamefont {Hollmann}},
  \ and\ \bibinfo {author} {\bibfnamefont {R.}~\bibnamefont {W\"ordenweber}},\
  }\href {\doibase http://dx.doi.org/10.1063/1.2216819} {\bibfield  {journal}
  {\bibinfo  {journal} {J. Appl. Phys.}\ }\textbf {\bibinfo {volume} {100}},\
  \bibinfo {eid} {023913} (\bibinfo {year} {2006}),\
  http://dx.doi.org/10.1063/1.2216819}\BibitemShut {NoStop}%
\bibitem [{\citenamefont {Gonz\'alez}\ \emph {et~al.}(2007)\citenamefont
  {Gonz\'alez}, \citenamefont {Hollmann},\ and\ \citenamefont
  {W\"ordenweber}}]{Gon07jap}%
  \BibitemOpen
  \bibfield  {author} {\bibinfo {author} {\bibfnamefont {M.~P.}\ \bibnamefont
  {Gonz\'alez}}, \bibinfo {author} {\bibfnamefont {E.}~\bibnamefont
  {Hollmann}}, \ and\ \bibinfo {author} {\bibfnamefont {R.}~\bibnamefont
  {W\"ordenweber}},\ }\href {\doibase 10.1063/1.2781515} {\bibfield  {journal}
  {\bibinfo  {journal} {J. Appl. Phys.}\ }\textbf {\bibinfo {volume} {102}},\
  \bibinfo {eid} {063904} (\bibinfo {year} {2007})}\BibitemShut {NoStop}%
\bibitem [{\citenamefont {Soroka}\ \emph {et~al.}(2007)\citenamefont {Soroka},
  \citenamefont {Shklovskij},\ and\ \citenamefont {Huth}}]{Sor07prb}%
  \BibitemOpen
  \bibfield  {author} {\bibinfo {author} {\bibfnamefont {O.~K.}\ \bibnamefont
  {Soroka}}, \bibinfo {author} {\bibfnamefont {V.~A.}\ \bibnamefont
  {Shklovskij}}, \ and\ \bibinfo {author} {\bibfnamefont {M.}~\bibnamefont
  {Huth}},\ }\href {\doibase 10.1103/PhysRevB.76.014504} {\bibfield  {journal}
  {\bibinfo  {journal} {Phys. Rev. B}\ }\textbf {\bibinfo {volume} {76}},\
  \bibinfo {pages} {014504} (\bibinfo {year} {2007})}\BibitemShut {NoStop}%
\bibitem [{\citenamefont {Dobrovolskiy}\ \emph {et~al.}(2012)\citenamefont
  {Dobrovolskiy}, \citenamefont {Begun}, \citenamefont {Huth},\ and\
  \citenamefont {Shklovskij}}]{Dob12njp}%
  \BibitemOpen
  \bibfield  {author} {\bibinfo {author} {\bibfnamefont {O.~V.}\ \bibnamefont
  {Dobrovolskiy}}, \bibinfo {author} {\bibfnamefont {E.}~\bibnamefont {Begun}},
  \bibinfo {author} {\bibfnamefont {M.}~\bibnamefont {Huth}}, \ and\ \bibinfo
  {author} {\bibfnamefont {V.~A.}\ \bibnamefont {Shklovskij}},\ }\href
  {http://stacks.iop.org/1367-2630/14/i=11/a=113027} {\bibfield  {journal}
  {\bibinfo  {journal} {New J. Phys.}\ }\textbf {\bibinfo {volume} {14}},\
  \bibinfo {pages} {113027} (\bibinfo {year} {2012})}\BibitemShut {NoStop}%
\bibitem [{\citenamefont {Dobrovolskiy}\ \emph {et~al.}(2010)\citenamefont
  {Dobrovolskiy}, \citenamefont {Huth},\ and\ \citenamefont
  {Shklovskij}}]{Dob10sst}%
  \BibitemOpen
  \bibfield  {author} {\bibinfo {author} {\bibfnamefont {O.~V.}\ \bibnamefont
  {Dobrovolskiy}}, \bibinfo {author} {\bibfnamefont {M.}~\bibnamefont {Huth}},
  \ and\ \bibinfo {author} {\bibfnamefont {V.~A.}\ \bibnamefont {Shklovskij}},\
  }\href {\doibase doi:10.1088/0953-2048/23/12/125014} {\bibfield  {journal}
  {\bibinfo  {journal} {Supercond. Sci. Technol.}\ }\textbf {\bibinfo {volume}
  {23}},\ \bibinfo {pages} {125014} (\bibinfo {year} {2010})}\BibitemShut
  {NoStop}%
\bibitem [{\citenamefont {Dobrovolskiy}\ \emph {et~al.}(2011)\citenamefont
  {Dobrovolskiy}, \citenamefont {Begun}, \citenamefont {Huth}, \citenamefont
  {Shklovskij},\ and\ \citenamefont {Tsindlekht}}]{Dob11pcs}%
  \BibitemOpen
  \bibfield  {author} {\bibinfo {author} {\bibfnamefont {O.~V.}\ \bibnamefont
  {Dobrovolskiy}}, \bibinfo {author} {\bibfnamefont {E.}~\bibnamefont {Begun}},
  \bibinfo {author} {\bibfnamefont {M.}~\bibnamefont {Huth}}, \bibinfo {author}
  {\bibfnamefont {V.~A.}\ \bibnamefont {Shklovskij}}, \ and\ \bibinfo {author}
  {\bibfnamefont {M.~I.}\ \bibnamefont {Tsindlekht}},\ }\href {\doibase
  10.1016/j.physc.2011.05.245} {\bibfield  {journal} {\bibinfo  {journal}
  {Physica C}\ }\textbf {\bibinfo {volume} {471}},\ \bibinfo {pages} {449}
  (\bibinfo {year} {2011})}\BibitemShut {NoStop}%
\bibitem [{\citenamefont {Silhanek}\ \emph {et~al.}(2010)\citenamefont
  {Silhanek}, \citenamefont {Van~de Vondel},\ and\ \citenamefont
  {Moshchalkov}}]{Sil10inb}%
  \BibitemOpen
  \bibfield  {author} {\bibinfo {author} {\bibfnamefont {A.~V.}\ \bibnamefont
  {Silhanek}}, \bibinfo {author} {\bibfnamefont {J.}~\bibnamefont {Van~de
  Vondel}}, \ and\ \bibinfo {author} {\bibfnamefont {V.~V.}\ \bibnamefont
  {Moshchalkov}},\ }\enquote {\bibinfo {title} {Guided vortex motion and vortex
  ratchets in nanostructured superconductors},}\ in\ \href@noop {} {\emph
  {\bibinfo {booktitle} {Nanoscience and Engineering in Superconductivity}}}\
  (\bibinfo  {publisher} {Springer-Verlag, Berlin Heidelberg},\ \bibinfo {year}
  {2010})\ Chap.~\bibinfo {chapter} {1}, pp.\ \bibinfo {pages}
  {1--24}\BibitemShut {NoStop}%
\bibitem [{\citenamefont {Dobrovolskiy}(2015{\natexlab{a}})}]{Dob15arx}%
  \BibitemOpen
  \bibfield  {author} {\bibinfo {author} {\bibfnamefont {O.~V.}\ \bibnamefont
  {Dobrovolskiy}},\ }\href@noop {} {\enquote {\bibinfo {title} {Abrikosov
  fluxonics in washboard nanolandscapes},}\ } (\bibinfo {year}
  {2015}{\natexlab{a}})\BibitemShut {NoStop}%
\bibitem [{\citenamefont {Kopelevich}\ \emph {et~al.}(1989)\citenamefont
  {Kopelevich}, \citenamefont {Lemanov}, \citenamefont {Sonin},\ and\
  \citenamefont {Kholkin}}]{Kop89etp}%
  \BibitemOpen
  \bibfield  {author} {\bibinfo {author} {\bibfnamefont {Y.~V.}\ \bibnamefont
  {Kopelevich}}, \bibinfo {author} {\bibfnamefont {V.}~\bibnamefont {Lemanov}},
  \bibinfo {author} {\bibfnamefont {E.~B.}\ \bibnamefont {Sonin}}, \ and\
  \bibinfo {author} {\bibfnamefont {A.~L.}\ \bibnamefont {Kholkin}},\
  }\href@noop {} {\bibfield  {journal} {\bibinfo  {journal} {JETP Lett.}\
  }\textbf {\bibinfo {volume} {50}},\ \bibinfo {pages} {212} (\bibinfo {year}
  {1989})}\BibitemShut {NoStop}%
\bibitem [{\citenamefont {Shklovskij}\ \emph {et~al.}(1999)\citenamefont
  {Shklovskij}, \citenamefont {Soroka},\ and\ \citenamefont
  {Soroka}}]{Shk99etp}%
  \BibitemOpen
  \bibfield  {author} {\bibinfo {author} {\bibfnamefont {V.~A.}\ \bibnamefont
  {Shklovskij}}, \bibinfo {author} {\bibfnamefont {A.~K.}\ \bibnamefont
  {Soroka}}, \ and\ \bibinfo {author} {\bibfnamefont {A.~A.}\ \bibnamefont
  {Soroka}},\ }\href@noop {} {\bibfield  {journal} {\bibinfo  {journal} {J.
  Exp. Theor. Phys.}\ }\textbf {\bibinfo {volume} {89}},\ \bibinfo {pages}
  {1138} (\bibinfo {year} {1999})}\BibitemShut {NoStop}%
\bibitem [{\citenamefont {Shklovskij}(2003)}]{Shk03ltp}%
  \BibitemOpen
  \bibfield  {author} {\bibinfo {author} {\bibfnamefont {V.~A.}\ \bibnamefont
  {Shklovskij}},\ }\href {\doibase 10.1023/A:1023499615213} {\bibfield
  {journal} {\bibinfo  {journal} {J. Low Temp. Phys.}\ }\textbf {\bibinfo
  {volume} {131}},\ \bibinfo {pages} {899} (\bibinfo {year}
  {2003})}\BibitemShut {NoStop}%
\bibitem [{\citenamefont {Shklovskij}(2005)}]{Shk05ltp}%
  \BibitemOpen
  \bibfield  {author} {\bibinfo {author} {\bibfnamefont {A.~V.}\ \bibnamefont
  {Shklovskij}},\ }\href {\doibase 10.1007/s10909-005-3932-1} {\bibfield
  {journal} {\bibinfo  {journal} {J. Low Temp. Phys.}\ }\textbf {\bibinfo
  {volume} {139}},\ \bibinfo {pages} {289} (\bibinfo {year}
  {2005})}\BibitemShut {NoStop}%
\bibitem [{\citenamefont {Shklovskij}\ and\ \citenamefont
  {Dobrovolskiy}(2006)}]{Shk06prb}%
  \BibitemOpen
  \bibfield  {author} {\bibinfo {author} {\bibfnamefont {V.~A.}\ \bibnamefont
  {Shklovskij}}\ and\ \bibinfo {author} {\bibfnamefont {O.~V.}\ \bibnamefont
  {Dobrovolskiy}},\ }\href {\doibase 10.1103/PhysRevB.74.104511} {\bibfield
  {journal} {\bibinfo  {journal} {Phys. Rev. B}\ }\textbf {\bibinfo {volume}
  {74}},\ \bibinfo {pages} {104511} (\bibinfo {year} {2006})}\BibitemShut
  {NoStop}%
\bibitem [{\citenamefont {Brandt}(1995)}]{Bra95rpp}%
  \BibitemOpen
  \bibfield  {author} {\bibinfo {author} {\bibfnamefont {E.~H.}\ \bibnamefont
  {Brandt}},\ }\href {http://stacks.iop.org/0034-4885/58/i=11/a=003} {\bibfield
   {journal} {\bibinfo  {journal} {Rep. Progr. Phys.}\ }\textbf {\bibinfo
  {volume} {58}},\ \bibinfo {pages} {1465} (\bibinfo {year}
  {1995})}\BibitemShut {NoStop}%
\bibitem [{\citenamefont {Vinokur}\ \emph {et~al.}(1993)\citenamefont
  {Vinokur}, \citenamefont {Geshkenbein}, \citenamefont {Feigel'man},\ and\
  \citenamefont {Blatter}}]{Vin93prl}%
  \BibitemOpen
  \bibfield  {author} {\bibinfo {author} {\bibfnamefont {V.~M.}\ \bibnamefont
  {Vinokur}}, \bibinfo {author} {\bibfnamefont {V.~B.}\ \bibnamefont
  {Geshkenbein}}, \bibinfo {author} {\bibfnamefont {M.~V.}\ \bibnamefont
  {Feigel'man}}, \ and\ \bibinfo {author} {\bibfnamefont {G.}~\bibnamefont
  {Blatter}},\ }\href {\doibase 10.1103/PhysRevLett.71.1242} {\bibfield
  {journal} {\bibinfo  {journal} {Phys. Rev. Lett.}\ }\textbf {\bibinfo
  {volume} {71}},\ \bibinfo {pages} {1242} (\bibinfo {year}
  {1993})}\BibitemShut {NoStop}%
\bibitem [{\citenamefont {Wang}\ \emph {et~al.}(1994)\citenamefont {Wang},
  \citenamefont {Dong},\ and\ \citenamefont {Ting}}]{Wan94prl}%
  \BibitemOpen
  \bibfield  {author} {\bibinfo {author} {\bibfnamefont {Z.~D.}\ \bibnamefont
  {Wang}}, \bibinfo {author} {\bibfnamefont {J.}~\bibnamefont {Dong}}, \ and\
  \bibinfo {author} {\bibfnamefont {C.~S.}\ \bibnamefont {Ting}},\ }\href
  {\doibase 10.1103/PhysRevLett.72.3875} {\bibfield  {journal} {\bibinfo
  {journal} {Phys. Rev. Lett.}\ }\textbf {\bibinfo {volume} {72}},\ \bibinfo
  {pages} {3875} (\bibinfo {year} {1994})}\BibitemShut {NoStop}%
\bibitem [{\citenamefont {Vasek}\ \emph {et~al.}(2004)\citenamefont {Vasek},
  \citenamefont {Shimakage},\ and\ \citenamefont {Wang}}]{Vas04pcs}%
  \BibitemOpen
  \bibfield  {author} {\bibinfo {author} {\bibfnamefont {P.}~\bibnamefont
  {Vasek}}, \bibinfo {author} {\bibfnamefont {H.}~\bibnamefont {Shimakage}}, \
  and\ \bibinfo {author} {\bibfnamefont {Z.}~\bibnamefont {Wang}},\ }\href
  {\doibase http://dx.doi.org/10.1016/j.physc.2004.07.011} {\bibfield
  {journal} {\bibinfo  {journal} {Physica C}\ }\textbf {\bibinfo {volume}
  {411}},\ \bibinfo {pages} {164 } (\bibinfo {year} {2004})}\BibitemShut
  {NoStop}%
\bibitem [{\citenamefont {Huth}\ \emph {et~al.}(2002)\citenamefont {Huth},
  \citenamefont {Ritley}, \citenamefont {Oster}, \citenamefont {Dosch},\ and\
  \citenamefont {Adrian}}]{Hut02afm}%
  \BibitemOpen
  \bibfield  {author} {\bibinfo {author} {\bibfnamefont {M.}~\bibnamefont
  {Huth}}, \bibinfo {author} {\bibfnamefont {K.}~\bibnamefont {Ritley}},
  \bibinfo {author} {\bibfnamefont {J.}~\bibnamefont {Oster}}, \bibinfo
  {author} {\bibfnamefont {H.}~\bibnamefont {Dosch}}, \ and\ \bibinfo {author}
  {\bibfnamefont {H.}~\bibnamefont {Adrian}},\ }\href {\doibase
  10.1002/1616-3028(20020517)12:5<333::AID-ADFM333>3.0.CO;2-C} {\bibfield
  {journal} {\bibinfo  {journal} {Adv. Func. Mat.}\ }\textbf {\bibinfo {volume}
  {12}},\ \bibinfo {pages} {333} (\bibinfo {year} {2002})}\BibitemShut
  {NoStop}%
\bibitem [{\citenamefont {Dobrovolskiy}\ and\ \citenamefont
  {Huth}(2012)}]{Dob12tsf}%
  \BibitemOpen
  \bibfield  {author} {\bibinfo {author} {\bibfnamefont {O.~V.}\ \bibnamefont
  {Dobrovolskiy}}\ and\ \bibinfo {author} {\bibfnamefont {M.}~\bibnamefont
  {Huth}},\ }\href {\doibase 10.1016/j.tsf.2012.04.083} {\bibfield  {journal}
  {\bibinfo  {journal} {Thin Solid Films}\ }\textbf {\bibinfo {volume} {520}},\
  \bibinfo {pages} {5985} (\bibinfo {year} {2012})}\BibitemShut {NoStop}%
\bibitem [{\citenamefont {Lu}\ \emph {et~al.}(2007)\citenamefont {Lu},
  \citenamefont {Reichhardt},\ and\ \citenamefont {Reichhardt}}]{Luq07prb}%
  \BibitemOpen
  \bibfield  {author} {\bibinfo {author} {\bibfnamefont {Q.}~\bibnamefont
  {Lu}}, \bibinfo {author} {\bibfnamefont {C.~J.~O.}\ \bibnamefont
  {Reichhardt}}, \ and\ \bibinfo {author} {\bibfnamefont {C.}~\bibnamefont
  {Reichhardt}},\ }\href {\doibase 10.1103/PhysRevB.75.054502} {\bibfield
  {journal} {\bibinfo  {journal} {Phys. Rev. B}\ }\textbf {\bibinfo {volume}
  {75}},\ \bibinfo {pages} {054502} (\bibinfo {year} {2007})}\BibitemShut
  {NoStop}%
\bibitem [{\citenamefont {Dobrovolskiy}\ \emph
  {et~al.}(2015{\natexlab{a}})\citenamefont {Dobrovolskiy}, \citenamefont
  {Franke},\ and\ \citenamefont {Huth}}]{Dob15mst}%
  \BibitemOpen
  \bibfield  {author} {\bibinfo {author} {\bibfnamefont {O.~V.}\ \bibnamefont
  {Dobrovolskiy}}, \bibinfo {author} {\bibfnamefont {J.}~\bibnamefont
  {Franke}}, \ and\ \bibinfo {author} {\bibfnamefont {M.}~\bibnamefont
  {Huth}},\ }\href {http://stacks.iop.org/0957-0233/26/i=3/a=035502} {\bibfield
   {journal} {\bibinfo  {journal} {Meas. Sci. Technol.}\ }\textbf {\bibinfo
  {volume} {26}},\ \bibinfo {pages} {035502} (\bibinfo {year}
  {2015}{\natexlab{a}})}\BibitemShut {NoStop}%
\bibitem [{\citenamefont {Dobrovolskiy}(2015{\natexlab{b}})}]{Dob15snm}%
  \BibitemOpen
  \bibfield  {author} {\bibinfo {author} {\bibfnamefont {O.~V.}\ \bibnamefont
  {Dobrovolskiy}},\ }\href {\doibase
  http://dx.doi.org/10.1007/s10948-014-2664-3} {\bibfield  {journal} {\bibinfo
  {journal} {J. Supercond. Nov. Magnet.}\ }\textbf {\bibinfo {volume} {28}},\
  \bibinfo {pages} {469} (\bibinfo {year} {2015}{\natexlab{b}})}\BibitemShut
  {NoStop}%
\bibitem [{\citenamefont {Dobrovolskiy}\ and\ \citenamefont
  {Huth}(2015)}]{Dob15apl}%
  \BibitemOpen
  \bibfield  {author} {\bibinfo {author} {\bibfnamefont {O.~V.}\ \bibnamefont
  {Dobrovolskiy}}\ and\ \bibinfo {author} {\bibfnamefont {M.}~\bibnamefont
  {Huth}},\ }\href {\doibase http://dx.doi.org/10.1063/1.4917229} {\bibfield
  {journal} {\bibinfo  {journal} {Appl. Phys. Lett.}\ }\textbf {\bibinfo
  {volume} {106}},\ \bibinfo {pages} {142601} (\bibinfo {year}
  {2015})}\BibitemShut {NoStop}%
\bibitem [{\citenamefont {Dobrovolskiy}\ \emph
  {et~al.}(2015{\natexlab{b}})\citenamefont {Dobrovolskiy}, \citenamefont
  {Huth},\ and\ \citenamefont {Shklovskij}}]{Dob15met}%
  \BibitemOpen
  \bibfield  {author} {\bibinfo {author} {\bibfnamefont {O.~V.}\ \bibnamefont
  {Dobrovolskiy}}, \bibinfo {author} {\bibfnamefont {M.}~\bibnamefont {Huth}},
  \ and\ \bibinfo {author} {\bibfnamefont {V.~A.}\ \bibnamefont {Shklovskij}},\
  }\href {\doibase http://dx.doi.org/10.1063/1.4934487} {\bibfield  {journal}
  {\bibinfo  {journal} {Appl. Phys. Lett.}\ }\textbf {\bibinfo {volume}
  {107}},\ \bibinfo {pages} {162603} (\bibinfo {year}
  {2015}{\natexlab{b}})}\BibitemShut {NoStop}%
\bibitem [{\citenamefont {Silva}\ \emph {et~al.}(2017)\citenamefont {Silva},
  \citenamefont {Pompeo},\ and\ \citenamefont {Dobrovolskiy}}]{Sil17inb}%
  \BibitemOpen
  \bibfield  {author} {\bibinfo {author} {\bibfnamefont {E.}~\bibnamefont
  {Silva}}, \bibinfo {author} {\bibfnamefont {N.}~\bibnamefont {Pompeo}}, \
  and\ \bibinfo {author} {\bibfnamefont {O.}~\bibnamefont {Dobrovolskiy}},\
  }\enquote {\bibinfo {title} {Vortices at microwaves},}\ in\ \href@noop {}
  {\emph {\bibinfo {booktitle} {Superconductors at the Nanoscale: From Basic
  Research to Applications}}},\ \bibinfo {editor} {edited by\ \bibinfo {editor}
  {\bibfnamefont {R.}~\bibnamefont {W\"ordenweber}}, \bibinfo {editor}
  {\bibfnamefont {V.}~\bibnamefont {Moshchalkov}}, \bibinfo {editor}
  {\bibfnamefont {S.}~\bibnamefont {Bending}}, \ and\ \bibinfo {editor}
  {\bibfnamefont {F.}~\bibnamefont {Tafuri}}}\ (\bibinfo  {publisher} {Walter
  De Gruyter Inc., Berlin},\ \bibinfo {year} {2017})\ Chap.~\bibinfo {chapter}
  {18}\BibitemShut {NoStop}%
\bibitem [{\citenamefont {Koshelev}\ and\ \citenamefont
  {Vinokur}(1994)}]{Kos94prl}%
  \BibitemOpen
  \bibfield  {author} {\bibinfo {author} {\bibfnamefont {A.~E.}\ \bibnamefont
  {Koshelev}}\ and\ \bibinfo {author} {\bibfnamefont {V.~M.}\ \bibnamefont
  {Vinokur}},\ }\href {\doibase 10.1103/PhysRevLett.73.3580} {\bibfield
  {journal} {\bibinfo  {journal} {Phys. Rev. Lett.}\ }\textbf {\bibinfo
  {volume} {73}},\ \bibinfo {pages} {3580} (\bibinfo {year}
  {1994})}\BibitemShut {NoStop}%
\bibitem [{\citenamefont {Giamarchi}\ and\ \citenamefont
  {Le~Doussal}(1996)}]{Gia96prl}%
  \BibitemOpen
  \bibfield  {author} {\bibinfo {author} {\bibfnamefont {T.}~\bibnamefont
  {Giamarchi}}\ and\ \bibinfo {author} {\bibfnamefont {P.}~\bibnamefont
  {Le~Doussal}},\ }\href {\doibase 10.1103/PhysRevLett.76.3408} {\bibfield
  {journal} {\bibinfo  {journal} {Phys. Rev. Lett.}\ }\textbf {\bibinfo
  {volume} {76}},\ \bibinfo {pages} {3408} (\bibinfo {year}
  {1996})}\BibitemShut {NoStop}%
\end{thebibliography}

%

\end{document}